\documentclass[pra,aps,twocolumn,groupedaddress,floats,showpacs,final,superscriptaddress]{revtex4-1}
\usepackage{dcolumn}
\usepackage{bm}
\usepackage{color}
\usepackage{mathtools}
\usepackage{amsmath,amsfonts,amssymb,bm,ulem}

\usepackage{graphicx}
\usepackage{epsfig}
\usepackage{psfrag}
\usepackage{color}
\usepackage[pdftex]{hyperref} 

\newcommand{\rsub}{} 
\newcommand{\bl}{} 
\newcommand{\Vr}{\mathcal{V}}
\newcommand{\be}{\begin{equation}}
\newcommand{\ee}{\end{equation}}
\newcommand{\Nmax}{N_{\rm max}}
\newcommand{\ra}{\rangle}
\newcommand{\la}{\langle}

\begin{document}


\title{High-precision numerical solution of the Fermi polaron problem
\\and large-order behavior of its diagrammatic series}

\author{Kris Van Houcke}
\affiliation{Laboratoire de Physique de l'Ecole Normale Sup\'erieure, ENS, Universit\'e PSL, CNRS, Sorbonne Universit\'e, Universit\'e de Paris,
24 rue Lhomond,
75005 Paris, France}
\author{F{\'e}lix Werner}
\affiliation{Laboratoire Kastler Brossel, Ecole Normale Sup\'erieure, Universit\'e PSL, CNRS, Sorbonne Universit\'e, Coll\`ege de France, 
24 rue Lhomond,
75005 Paris, France}
\author{Riccardo Rossi}
\affiliation{Center for Computational Quantum Physics, Flatiron Institute, 162 Fifth Avenue, New York, NY 10010, USA}

\date{\today}

\begin{abstract}
We introduce a simple 
determinant diagrammatic Monte Carlo algorithm to 
compute
the ground-state properties
of a particle 
interacting with a Fermi sea 
 through a zero-range interaction.
The fermionic sign does not cause any fundamental problem when going to high diagram orders,
and we
reach order $N=30$.
The data reveal that the diagrammatic series diverges exponentially as $(-1/R)^{N}$
with a radius of convergence $R<1$. 
Furthermore, 
on the polaron side of the polaron-dimeron transition, the value of $R$ is determined by a special class of three-body diagrams, corresponding to repeated scattering of the impurity between two  particles of the Fermi sea.
A power-counting argument explains why finite $R$ is possible
for zero-range interactions in three dimensions. 
Resumming the divergent series through a conformal mapping
yields
the polaron energy with record accuracy.

\end{abstract}


\maketitle

The  Fermi polaron is a quasi-particle that emerges when a 
mobile impurity 
is coupled 
through a short-range interaction
to a single-component ideal 
Fermi gas~\cite{Massignan_2014}. 
Its energy, mass and quasi-particle residue are renormalized since the bare particle is dressed by particle-hole excitations of the Fermi sea. 
Experimental studies with cold atomic gases~\cite{ZwierleinPolaron,nascimbene2009pol,polaron_Grimm,polaron_Kohl,GrimmPolaronDecoherence,GrimmUltrafast,LENSRepulsivePolaron,Boiling_MIT,FollingMultiorbPolaron} 
raise a considerable theoretical
interest in this system. 
While exact analytical results 
can be obtained in one dimension~\cite{ZvonarevFlutter,BurovskiImpurityRelaxation,Cui1DPolaron,ZvonarevImpurityInjection}
most works 
in higher dimensions rely on approximate treatments of the strongly correlated many-body problem~\cite{chevy2006upa,Combescot2007,Mora2009,ZwergerPunk,CGL2009,MassignanBruunRepulsive2011,MassignanBruunRepulsive2013,ParishHusePolaron,TrefzgerCastinEPL,QiZhaiPolaronNarrow,HuPolaronT,ParishPolaron2D,DemlerPolaron2D,ParishLevinsenRepulsivePolaron2D,ParishLevinsenPolaron2D,ParishDyna,ParishDynaT}.
It is also possible to solve the equilibrium problem in three dimensions in an unbiased way,
as demonstrated by Prokof'ev and Svistunov by means of a diagrammatic Monte Carlo (DiagMC) algorithm~\cite{ProkofevSvistunovPolaronShort,ProkofevSvistunovPolaronLong}.
This study 
established an important benchmark for the Fermi polaron quasi-particle properties, 
and showed
that beyond a critical strength
of the (attractive) interaction,
 the polaron becomes unstable
and a dimeron --a bosonic quasiparticle composed of the impurity bound with one fermion, dressed with particle-hole excitations-- is formed instead in the ground state. 
The work of Prokof'ev and Svistunov was extended to different observables~\cite{Vlietinck13,Goulko_Dark}, two dimensions~\cite{Vlietinck_2D,Pollet_polaron_2D}, and different impurity masses~\cite{polaron_DiagMC_Lode}.

In DiagMC algorithms, 
the sum over Feynman diagram topologies is done by stochastic sampling.
In contrast, in
{\it determinant} diagrammatic Monte Carlo,
one
sums exactly over all topologies at each Monte Carlo step,
using the fact that
(for a given set of positions and imaginary times of the vertices)
 the sum of all topologies,
including disconnected ones,
can be expressed as a determinant.
This approach, also known as
continuous-time interaction-expansion Quantum Monte~Carlo,
was
introduced for quantum impurity models~\footnote{These quantum impurity models ({\it e.g.} the Anderson model)
are fundamentally different from polaron models:
In the former, the impurity is a special location where the bath-particles can interact,
while in the latter, the impurity is a mobile particle which interacts with the bath particles.} in Refs.~\cite{Rubtsov2003,Rubtsov2004},
and is widely used as impurity solver within Dynamical Mean Field Theory~\cite{Rubtsov2005,GullRevueImpurityQMC}.
This approach
also used for direct evaluation of the many-body diagrammatic series for the Hubbard model~\cite{Burovski2004,KozikBurovskiNeel} 
including for the unitary Fermi gas by extrapolating to zero particles per site~\cite{Burovski2004,KozikBurovskiNeel,zhenyaPRL,zhenyaNJP,GoulkoPRA,GoulkoWingate2015},
and for electron-phonon models in 1D~\cite{AssaadDDMC2012,AssaadDDMC2015,AssaadDDMC2016}.
These works mostly restricted to special sign-free cases (repulsive half-filled or attractive unpolarized Hubbard model, dispersionless phonons),
because generically,
 disconnected topologies 
(which do not contribute to the final result for local quantities) cause a sign problem leading to an exponential increase of CPU-time with system size.

In DiagMC simulations of many-body systems,
system size does not play any essential role,
since one restricts the sampling to connected topologies~\cite{VanHoucke1short}. 
More generally, it is easy in DiagMC to discard certain topologies:
one-particle reducible diagrams, as required to compute the self-energy;
diagrams containing bubbles, as required to work with dressed vertices containing all ladder diagrams;
or two-particle reducible diagrams, as required to compute skeleton series built with fully dressed propagators.
This 
led to
results directly in the thermodynamic limit for various many-body problems, {\it e.g.},
the Hubbard model in doped repulsive~\cite{KozikVanHouckeEPL,kozik_pseudogap,JohanInfiniteU}
or polarized attractive regimes~\cite{GukelbergerFFLO},
the unitary Fermi gas directly in the continuous space zero-range limit~\cite{VanHouckeEOS,RossiEOS,RossiContact},
electron-phonon models without~\cite{MishchenkoProkofevPRL2014}
or with additional Coulomb interaction~\cite{IgorCoulombPhonon},
graphene~\cite{IgorDirac},
Weyl semimetals~\cite{CarlstromWeyl},
topological phases of the Haldane-Hubbard-Coulomb model~\cite{IgorHaldane},
or frustrated spins~\cite{KulaginPRL,HuangPyro,WangCaiSpins}.

Recently a new type of determinant diagrammatic Monte Carlo algorithm was introduced for the Fermi-Hubbard model, where disconnected or reducible topologies are removed by recursive formulas~\cite{RossiCDet,SimkovicCDet,MoutenetCDet,RossiSelf},
an approach we will refer to as CDet, for connected determinant.
The recursive formulas require a number of operations that scales exponentially with the diagram order, 
and this is done at each Monte~Carlo step.
Still, this pays off because it takes better advantage of massive cancellations between topologies 
---the average sign decreases only exponentially with the order for CDet,
while for DiagMC it decreases factorially, because a factorial number of topologies needs to be sampled stochastically.
The CDet approach was already used to study the metal-to-insulator crossover in the half-filled Hubbard model~\cite{SimkovicCrossover,KimCrossover}.

The general idea of using determinants and paying an exponential price at each Monte~Carlo update in order to remove disconnected diagrams was actually first introduced for the real-time evolution of the Anderson impurity model after switching on the interaction,
where no recursive formula is needed as it suffices to sum over Keldysh indices; the elimination of disconnected diagrams made it possible to reach the long-time limit, {\it i.e.} the steady-state regime, and to obtain previously unaccessible results~\cite{Profumo,Bertrand1,Bertrand2}.
A new way of
combining determinants with recursive relations was recently found to be more efficient than CDet
in the hybridisation-expansion approach
 to the Anderson impurity model in real time,
while CDet remains faster for the interaction expansion~\cite{CohenInchwormDet}.
The strategy of summing over all connected topologies at each Monte~Carlo update was also found to be useful for the electron gas,
even though no determinants were used
and the number of operations for summing over topologies was factorial in diagram order~\cite{KunHauleEG}.

In this article, we introduce a determinant diagrammatic Monte Carlo algorithm for solving the Fermi-polaron model, based on several simple observations.
First, 
summing up topologies using a determinant does not generate any 
disconnected diagram because such diagrams do not exist for Fermi-polaron problems~\footnote{In an independent work, this was also observed and used to evaluate bare series for a repulsive square-well potential (L.~Pollet, talk at the {\it Diagrammatic Monte Carlo Workshop}, Flatiron Institute, New York, July 2019).}. 
Second, a recursive formula 
allows 
to remove one-particle reducible diagrams and thereby calculate the polaron 
self-energy. 
Third, particle-particle bubbles are easily removed by setting to zero appropriate matrix elements before taking the determinants,
which allows to use dressed vertices including all ladder diagrams,
and thus to work directly with zero-range interactions in continuous space.

The new algorithm,
which we will refer to as PDet, for polaron determinant,
 is easier to program than DiagMC and allows to go to much higher expansion orders. 
DiagMC simulations did not go beyond
order $12$ due to the factorial complexity of sampling a factorial number of diagrams~\cite{ProkofevSvistunovPolaronShort,Vlietinck13}. With
the PDet algorithm,
however, the scaling with diagram order is polynomial and we 
reach diagram order $30$.
The data reveal that the series is divergent for all values of the interaction strength,
and that the large-order behavior is essentially determined by particular scale-invariant diagrams corresponding to the three-body problem
as long as the polaron is the ground state.
Through a conformal mapping, the divergent series is converted into a convergent one.
As a first illustration we
compute
the polaron energy at unitarity with unprecedented level of precision and control.

\section{Model and diagrams}

We start by briefly reviewing the Fermi-polaron model and its diagrammatic formalism. 
The system can be viewed as a two-component Fermi gas
with only one spin-$\downarrow$ particle.
It is convenient to start from a lattice model 
of Hamiltonian
\begin{multline}
\hat{H}  =      \sum_{\mathbf{k}\in \mathcal{B}, \sigma = \uparrow, \downarrow} 
\frac{k^2}{2m}
 ~ \hat{c}^{\dagger}_{\mathbf{k} \sigma}
\hat{c}^{\phantom{\dagger}}_{\mathbf{k} \sigma}
\\ +  g_{0}    \sum_{\mathbf{r}} b^3 ~ \hat{\Psi}^{\dagger}_{\uparrow} (\mathbf{r})  \hat{\Psi}^{\dagger}_{\downarrow} (\mathbf{r}) \hat{\Psi}^{\phantom{\dagger}}_{\downarrow}(\mathbf{r}) \hat{\Psi}^{\phantom{\dagger}}_{\uparrow}(\mathbf{r})
\label{eq:ham}
\end{multline}
with  $\hat{\Psi}^{\phantom{\dagger}}_{\sigma}(\mathbf{r})$ and $ \hat{c}^{\phantom{\dagger}}_{\mathbf{k},\sigma}$ the field operators for
annihilating a spin-$\sigma$ fermion  in position and momentum space, respectively.
We set $\hbar=1$, and
 take the same mass $m$
for the fermions and the impurity. 
The components of the position  vector $\mathbf{r}$ are integer multiples of the lattice spacing $b$. 
Further, $g_0$ is the bare attractive interaction strength. The wave vectors $\mathbf{k}$ are in the first Brillouin zone $\mathcal{B} = ]-\pi/b, \pi/b]^3$. 
We consider zero temperature, 
so that the spin-$\uparrow$ particles form
a Fermi sea,
occupying states 
up to the Fermi energy $\varepsilon_F$ and Fermi momentum $k_F$.

A standard way to arrive at a diagrammatic series that is well-defined in the continuum limit (i.e. $b
\to 0^+$ and $g_0\to 0^-$ for a fixed scattering length $a$) is to first sum all ladder diagrams. This calculation gives a dressed interaction {\rsub 
line}
 $\Gamma^0$ (which {\rsub 
can be viewed as a partially dressed} bosonic pair propagator):
\begin{equation}
\includegraphics[width=0.9\columnwidth]{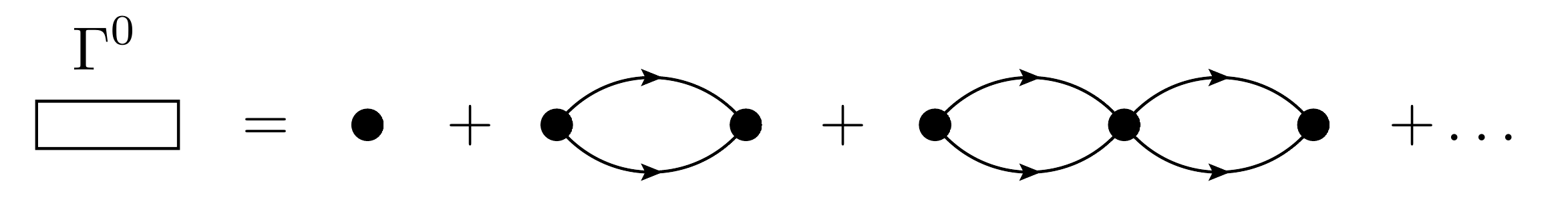} 
\end{equation}
where the black dot represents the bare  interaction $g_0$
(external lines are not shown). 
For more details and an explicit expression for $\Gamma^0$ we refer to Refs.~\cite{ProkofevSvistunovPolaronLong,Vlietinck13}. 
Then, Feynman diagrams can be built from
 the free single-particle propagators $G^{0}_{\sigma}$ and the pair propagators $\Gamma^0$.
Note that
at short {\rsub imaginary} time $\tau\to 0^+$ and short distance $r\lesssim\sqrt{\tau}$, the propagators  behave as 
 \begin{eqnarray}
 G^0_{\downarrow}(r, \tau)  & \sim &   \frac{1}{ \tau^{3/2}} ~e^{ - \frac{m}{2\tau} r^2}  \; , \label{eq:Gshort} \\
 \Gamma^0(r, \tau) & \sim &   \frac{1}{ \tau^2 }    e^{-\frac{m}{\tau}r^2}  \;. \label{eq:Gamshort} 
 \end{eqnarray}

\begin{figure}
\includegraphics[scale=0.38,width=0.99\columnwidth]{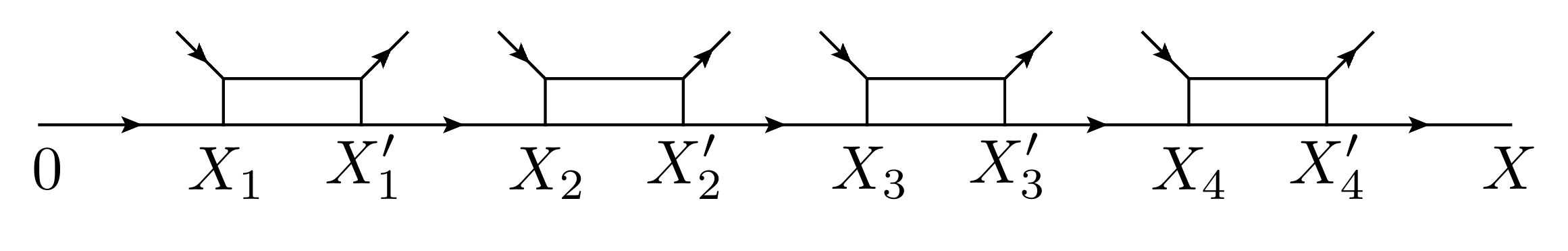}
\caption{Backbone structure of the fourth-order Feynman diagrams contributing to the {\rsub 
impurity propagator} $G(X)$. Each 
{\bl rectangle} 
is a dressed interaction {\rsub 
line} $\Gamma^0$ running from space-time point $X_i = (\mathbf{r}_i, \tau_i)$ to $X_i' = (\mathbf{r}_i', \tau_i')$. Each full line is a free $G^{0}_\downarrow$ propagator running from $X_i'$ to $X_{i+1}$, with $X_0'=0$ and $X_5=X$. 
All Feynman diagrams of order 4 are obtained by considering all possible connections of the spin-$\uparrow$ open ends with $G^0_\uparrow$ propagators, with the exclusion of ladder diagrams. All ladder diagrams have already been summed in each $\Gamma^0$ {\bl line}.
}
\label{fig:backbone}
\end{figure}

 Since the impurity propagates only forward in time, the number of possible topologies is greatly restricted compared to the case of a finite $\downarrow$ density (treated in Refs.~\onlinecite{VanHouckeEOS,VanHouckeLong}). 
{\rsub Consider the diagrams contributing to the impurity propagator
$G(r,\tau) := -\langle {\rm T}\,\psi_\downarrow({\bf r},\tau) \psi_\downarrow^\dagger({\bf 0},0) \rangle$.}
Each diagram of order $N$ (which contains $N$ interaction lines $\Gamma^0$) has a backbone created by the forward moving impurity, see Fig.~\ref{fig:backbone}. 
All allowed topologies are obtained by connecting the open ends indicated by the arrows with free propagators $G^{0}_{\uparrow}$.  There is one exception in order to avoid double counting: 
Particle-particle bubbles
\begin{equation}
\includegraphics[width=0.3\columnwidth]{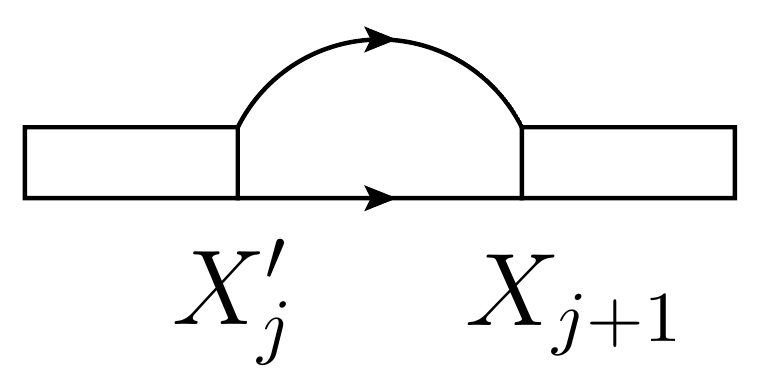} 
\label{eq:bubble}
\end{equation}
need to be omitted since they are already included in $\Gamma^0$. 
 
 \section{Monte~Carlo algorithm} \label{sec:algo}

We  turn to the 
description of the PDet algorithm. {\rsub 
As in Fig.~\ref{fig:backbone}, let $(X_1, X_1' ,\ldots, X_N, X_N') =: V_N$ be the internal space-time coordinates for a diagram of order $N$.}
The contribution of the backbone 
is given by
\begin{eqnarray}
& & B(V_N, X)  =   G^0_{\downarrow}(X_1) ~ \Gamma^0(X_1'-X_1) ~ G^0_{\downarrow}(X_2-X_1') \nonumber \\ & & ~\times ~\Gamma^0(X_2'-X_2)  \ldots ~ \Gamma^0(X_N'-X_N) ~ G^0_{\downarrow}(X-X'_N).
\end{eqnarray}
The sum of all possible ways  to close the open ends of the $\Gamma^0$-lines  with $G^0_{\uparrow}$-propagators
can be obtained by calculating a single determinant. The  $N$-th order contribution to the 
impurity propagator for external $X=(\mathbf{r}, \tau)$ is given by
\begin{eqnarray}
G_N(X) & =  & \int dX_1 \ldots  \int dX_N' ~ B(V_N, X)  
~S(V_N) \; ,
\label{eq:GNX}
\end{eqnarray}
with the notation $\int dX_i  = \int d\mathbf{r}_i \int_0^\infty d\tau_i$ and
where the sum of all possible connections by spin-up propagators is given by 
\begin{equation}
S(V_N) = {\rm det}[A(V_N)] \; ,
\end{equation}  
with the matrix elements of $A$ given by
\begin{equation}
  A_{i,j}=\begin{cases}
    G^0_{\uparrow}(X_i-X_j') & \text{if $i\neq j+1$} \; ,\\
    0 & \text{if $i=j+1$} \; .
  \end{cases}
\end{equation}
The zeros in the matrix $A$ ensure that ladder diagrams are not double counted, since it eliminates all particle-particle bubbles~(\ref{eq:bubble})
in the sum of possible connections.  
In the conventional diagrammatic Monte Carlo algorithm, all diagram topologies are sampled explicitly~\cite{ProkofevSvistunovPolaronShort,ProkofevSvistunovPolaronLong}. Here, their sum is given by just one determinant (for a given set of internal variables). This greatly simplifies the algorithm and the computer code. 
The price to pay is that one has to work in position representation, where the propagators are oscillating as a function of distance, which might be difficult to deal with numerically.  
The conventional DiagMC algorithm, on the other hand, works in momentum representation, where the propagators are sign definite as a function of momentum. 
In practice, the oscillating propagators do not turn out to be a limiting factor, and 
 the current algorithm is far superior to the conventional one. 

The PDet
algorithm 
stochastically performs the summation over the order $N$
and the multi-dimensional integral over internal variables
[see Eq.~(\ref{eq:GNX})] in order to calculate the contributions $G_N(X)$ up to some maximal order.
As usual in diagrammatic Monte~Carlo, the external variable $X$ is also sampled,
which allows to get the $X$-dependence from a histogram
(since we restrict here to zero external momentum, we only  histogram the $\tau$-dependence and not the $r$-dependence).
Accordingly, a configuration is given by $(V_N,X)$ and its weight is $W(V_N,X) = |B(V_N, X)S(V_N)|$.
An ergodic and efficient sampling
can be achieved through a few Monte~Carlo updates:

(i) \emph{Time shift.} Choose one backbone line at random. Let the space-time difference of this line be $\Delta X = (\Delta \mathbf{r}, \Delta \tau_{\rm old})$. Given $\Delta \mathbf{r}$, choose a new time 
difference $\Delta \tau_{\rm new}$ for this line proportional to the short time behavior given in Eq.~(\ref{eq:Gshort}) or Eq.~(\ref{eq:Gamshort}) in case the chosen backbone line is respectively a $G^0_{\downarrow}$ propagator or a $\Gamma^0$ propagator.

(ii) \emph{Position shift.} Choose one backbone line at random. Let the space-time difference of this line again be $\Delta X = (\Delta \mathbf{r}_{\rm old}, \Delta \tau)$. 
Given $\Delta \tau$, choose a new position
difference $\Delta \mathbf{r}_{\rm new}$ for this line according to a Gaussian with width $\Delta\tau/m$ in case of a $G^0_{\downarrow}$ propagator and with width $\Delta\tau/2m$ in case of $\Gamma^0$ propagator.  

(iii) \emph{Add.} Let the current configuration be $(V_N,X_{\rm old})$. The Add update will try to increase $N$ by one by simply adding one $\Gamma^0$-{\bl line} and one $G^0_{\downarrow}$ at the end of the $N$-th order backbone.  For the final configuration of order $N+1$ we take $X_{N+1} = X_{\rm old}$, while  a new $X_{N+1}'$ and a new total space-time diference $X_{\rm new}$ are chosen in the following way. First we choose the time difference $\Delta\tau_{N+1} = \tau'_{N+1}-\tau_{N+1}$ proportional to the short time behavior Eq.~(\ref{eq:Gamshort}) with $r = |\mathbf{r}_{N}'-\mathbf{r}_{N}|$.
Next the time difference $\Delta\tau_{\rm new} = \tau_{\rm new} - \tau'_{N+1}$ is chosen proportional to the short time behavior Eq.~(\ref{eq:Gshort}) with $r = |\mathbf{r}_{\rm old}-\mathbf{r}_{N}'|$. These choices of $r$ in the probability distribution should ensure that we do propose times which are typical.  Indeed, we observed that this gives good acceptance rates in practice. 
The new position difference $\Delta \mathbf{r}_{N+1} =  \mathbf{r}'_{N+1}-\mathbf{r}_{N+1}$ and $\Delta \mathbf{r}_{\rm new} = \mathbf{r}_{\rm new} - \mathbf{r}'_{N+1}$ are chosen from a Gaussian distribution with widths $\Delta\tau_{N+1}/2m$ and $\Delta\tau_{\rm new}/m$, respectively.  \

(iv) \emph{Remove}. This update is simply the inverse of the Add update.

As usual in diagrammatic Monte~Carlo, after a new configuration is proposed,
it is accepted or rejected using the Metropolis-Hastings {\bl rule},
the detailed balance being ensured separately for each complementary pair of updates (here, Add-Remove is a complementary pair, while each of the shift updates is self-complementary).
In our code we also use an additional update changing both time and position difference of a line, but this is not required for ergodicity. 
{\rsub To eliminate configurations with a very large weight,
we introduce a small lower cutoff on all position-differences along the backbone,
which we checked to induce a negligible systematic error.}
We also use standard reweighting procedures.
Namely, the function $W(V_N,X)$ is multiplied by an extra factor $C_N$ 
in order to spend a 
reasonable amount of simulation time at each order
(we simply choose to spend a constant number of Monte~Carlo steps per order).
 Moreover 
we use the
freedom to 
shift the impurity energy by a
free parameter $-\mu$, which amounts to 
exponential reweighting in time since each diagram of total time $\tau$ depends on $\mu$ through a simple factor $e^{\mu\tau}$~\footnote{We have set the  free parameter
$\mu/\varepsilon_F$ to -2.4, -1.2, -1.6, -2.8  for $1/(k_F a)$ = -1, 0, 0.5, 1 respectively.}.
To normalize the series,
we use the first order diagram which is calculated alternatively using Fourier transforms.

In order to calculate the polaron quasi-particle properties it is 
{\bl preferable} to calculate the self-energy $\Sigma(X)$ rather than $G(X)$.
To get the contribution $\Sigma_N(X)$  at order $N$ one needs to exclude all the {\bl one-particle} reducible diagrams. 
One has
\be
\Sigma_N(X) =    \int \!dX_1' \int \!dX_2  \ldots  \!\int \!dX_N ~ \tilde{B}(V_N) ~\tilde{S}(V_N)
\label{eq:Sigma_X}
\ee
with the backbone of the self-energy given by
\begin{eqnarray}
  & & \tilde{B}(V_N)  =   \Gamma^0(X_1') ~ G^0_{\downarrow}(X_2-X_1') \nonumber \\ & & ~\times ~\Gamma^0(X_2'-X_2)  \ldots ~ \Gamma^0(X-X_N) \; ,
  \label{eq:Btilde}
\end{eqnarray}
$\tilde{S}(V_N)$ is the sum of all connections with spin-up propagators 
that create one-particle irreducible diagrams,
and $X_1\equiv 0$ and $X_N'\equiv X$ in the set $V_N$ for the self-energy.  
Elimination of reducible diagrams is achieved by applying the following recursive relation at each step of the Monte Carlo process:
\begin{eqnarray}
  \tilde{S}(V_n) =  S(V_n) - \sum_{k=1}^{n-1} \tilde{S}(V_k) ~S(V_n \setminus V_k) \; ,
  \label{eq:Stilde}
\end{eqnarray}
for $n=1,\ldots N$. 
The computational cost for calculating $S(V_N)$ is $\mathcal{O}(N^3)$ since it requires just calculating one determinant.
The cost to calculate $\tilde{S}(V_N)$  is still polynomial.  If all the determinants in the recursive formula are calculated in a straightforward way without any special tricks, the calculation of $\tilde{S}(V_N)$
scales as $\sum_{n=1}^{N} \sum_{k=1}^n k^3 \sim N^5$.  The Monte~Carlo updates of the PDet algorithm for the self-energy are very similar to those of the algorithm for calculating $G$ described above. 

{\rsub 

Straightforward modifications of the above procedures
allow one to compute other quantities
---such as
the fully dressed pair propagator 
and the pair self-energy 
(denoted by $\Gamma$ and $\Pi$ in Ref.~\cite{Vlietinck13}),
which give access to the dimeron properties~\cite{ProkofevSvistunovPolaronLong,Vlietinck13}---
and to treat other fermionic polaron problems
---for example the bare expansion for 
a finite-range interaction.
}

\section{Large-order behavior}

In this section, we use the new algorithm 
to evaluate 
the diagrammatic series up to high order,
which 
then leads
us to investigate the asymptotic large-order behavior.

\subsection{Exponential divergence}\label{sec:radius}

\begin{figure}
\includegraphics[scale=0.38,width=0.99\columnwidth]{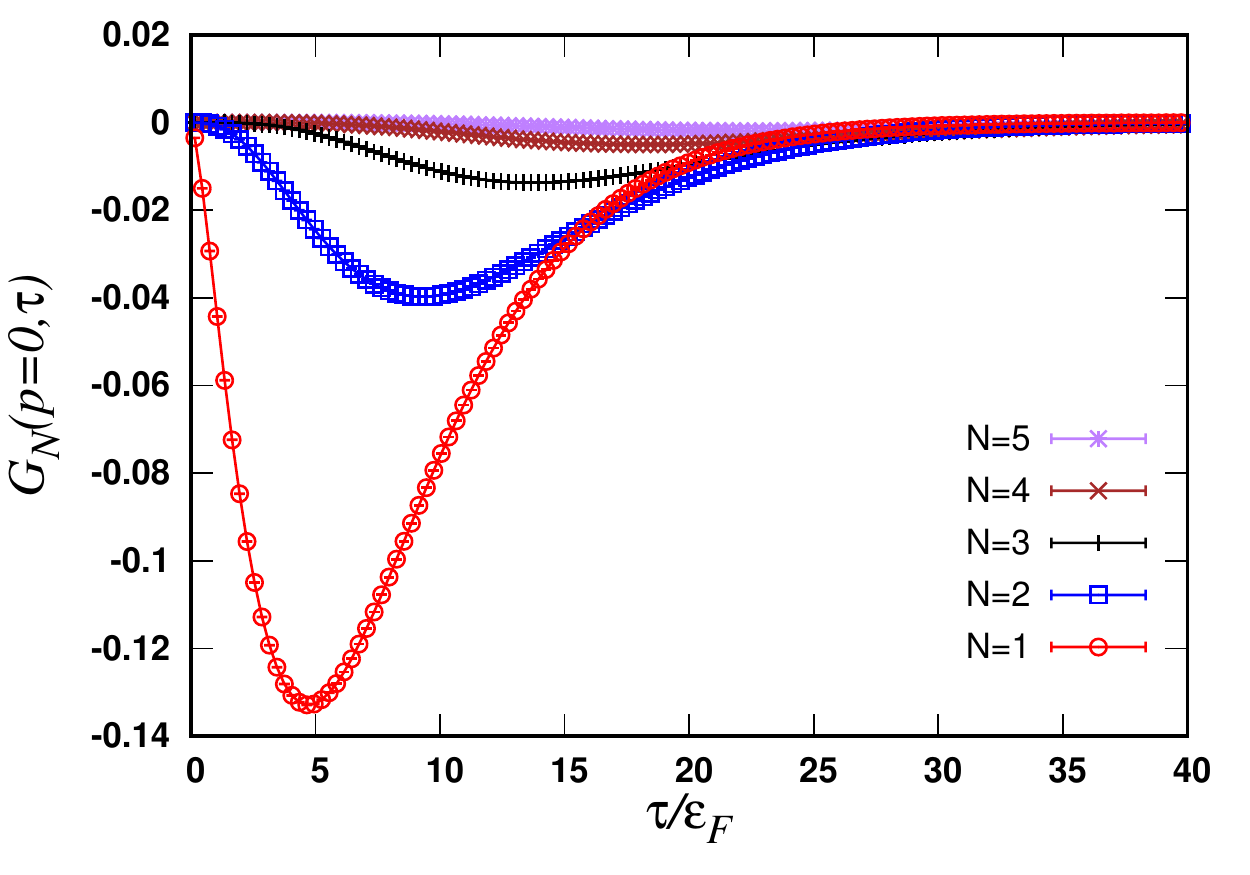}
\caption{Contribution to the 
impurity propagator
for the lowest orders $N$,
at zero momentum, as a function of imaginary time $\tau$. 
The interaction strength is $k_F a = \infty$.
}
\label{fig:G_N_tau5}
\end{figure}

\begin{figure}
\includegraphics[scale=0.38,width=0.99\columnwidth]{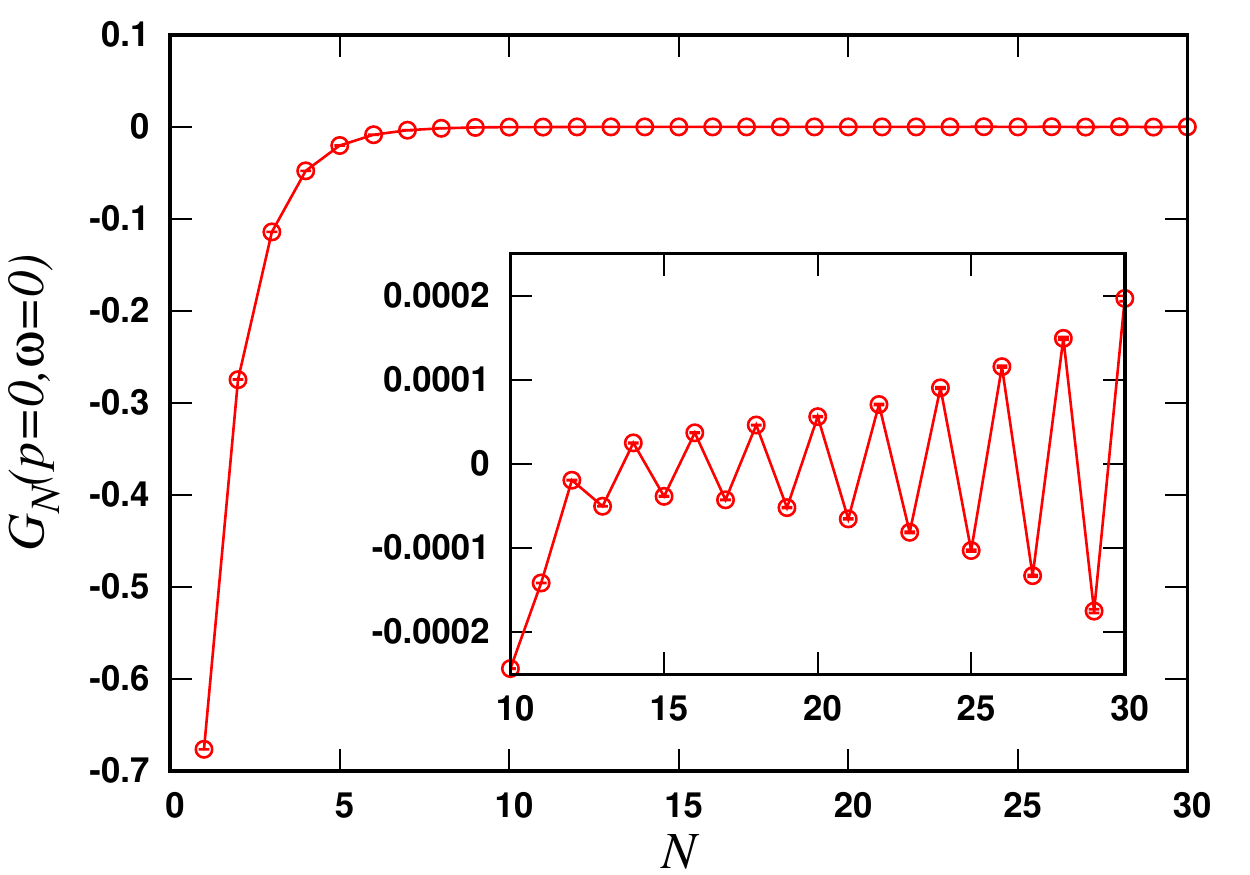}
\caption{The $N$-th order contribution $G_N$ to the 
impurity propagator 
at zero momentum and zero frequency for $k_F a = \infty$.  The data seems to be rapidly converging as a function of the order $N$. The zoom however shows an exponential and sign-alternating increase at high enough order. }
\label{fig:G_N_omega0}
\end{figure}

\begin{figure}
\includegraphics[scale=0.38,width=0.99\columnwidth]{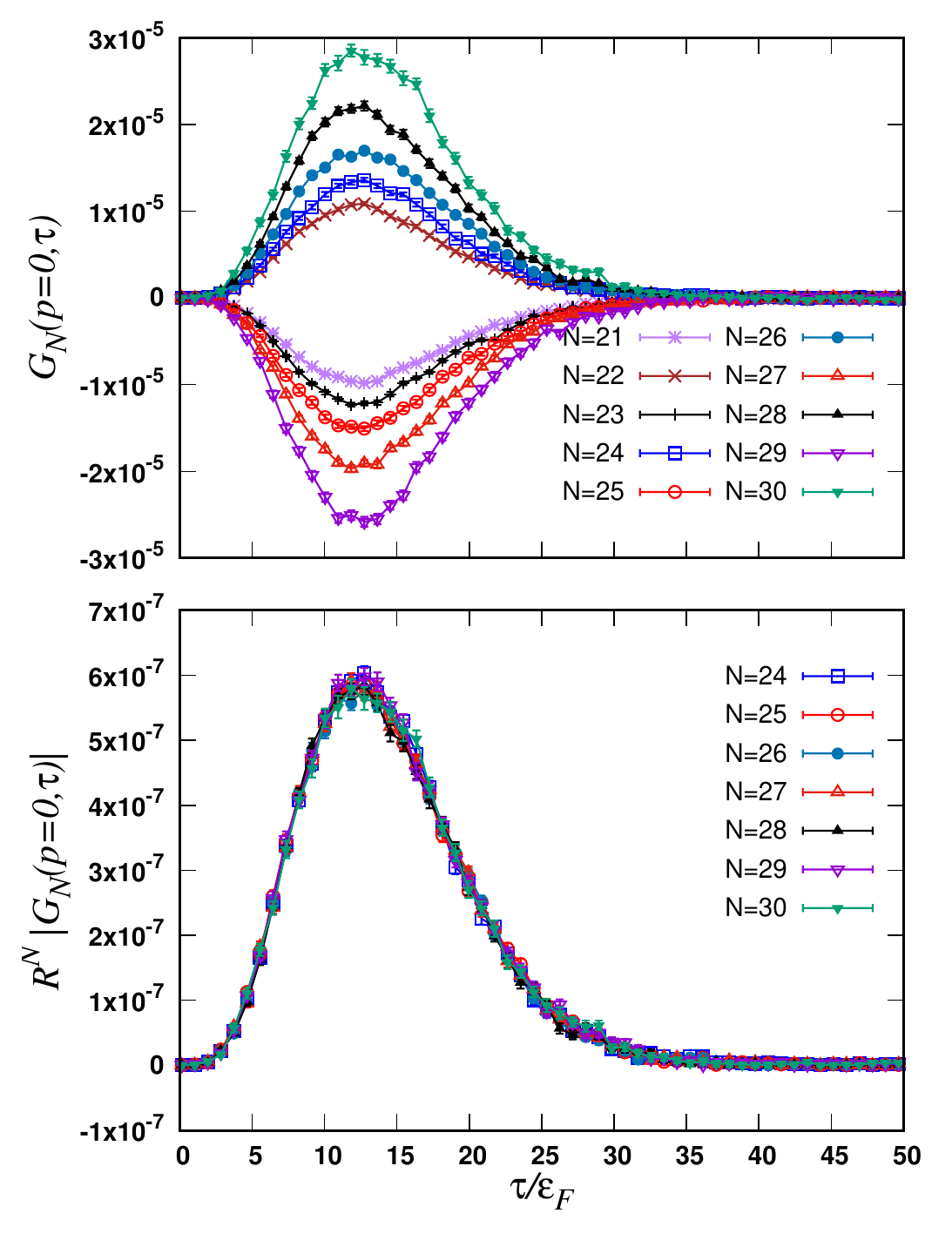}
\caption{The order-$N$ contribution to the 
impurity propagator
 $G(p=0,\tau)$ at zero momentum as a function of imaginary time $\tau$ for high orders. 
The interaction strength is $k_F a = \infty$. 
The lower panel shows $R^{N} |G_N(p=0,\tau)|$ as a function of imaginary time. The collapse of the data illustrates the behavior given in Eq.~(\ref{eq:RNF}) at large enough order $N$. 
}
\label{fig:G_N_tau30}
\end{figure}

We start by calculating the contributions to the impurity 
propagator for $k_F a = \infty$.  
In Fig.~\ref{fig:G_N_tau5} we show the order-$N$ contribution 
$G_N(\mathbf{p}=0,\tau)$,  for the lowest orders $N \leq 5$, as a function of imaginary time $\tau$. 
The series seems to  converge very rapidly as a function of $N$,
at least for small enough $\tau$.
But a closer inspection at high enough order shows that this is not the case.
In Fig.~\ref{fig:G_N_omega0}  we show $G_N(\mathbf{p}=0,\omega=0)  = \int_0^\infty d\tau\, G_N(\mathbf{p}=0,\tau)$ 
for orders $N$ up to 30:
Zooming on the high orders reveals 
that the 
series 
is actually diverging. 
The data is very well fit by an exponential sign-alternating law,
$G_N(p=0,\omega=0) = (-R)^{-N}$, with $R = 0.878(2)$
(taking a fitting range $N=24-30$).
Coming back to 
the imaginary-time dependence of $G_N(\mathbf{p}=0,\tau)$,
Fig.~\ref{fig:G_N_tau30}(a) shows that at large order $N$,
a peak  develops around $\tau/\varepsilon_F \approx 12$, with an amplitude that increases 
with $N$. 
More precisely,
Fig.~\ref{fig:G_N_tau30}(b) shows a data collapse
after multiplication by $(-R)^{N}$,
which indicates a large-order behavior
\begin{equation}
G_N(p=0,\tau)  \underset{N\to\infty}{\simeq} (-R)^{-N}  ~F(\tau) \;
\label{eq:RNF}
\end{equation}
were $F(\tau)$ is $N$-independent.

\begin{figure}
\includegraphics[scale=0.38,width=0.99\columnwidth]{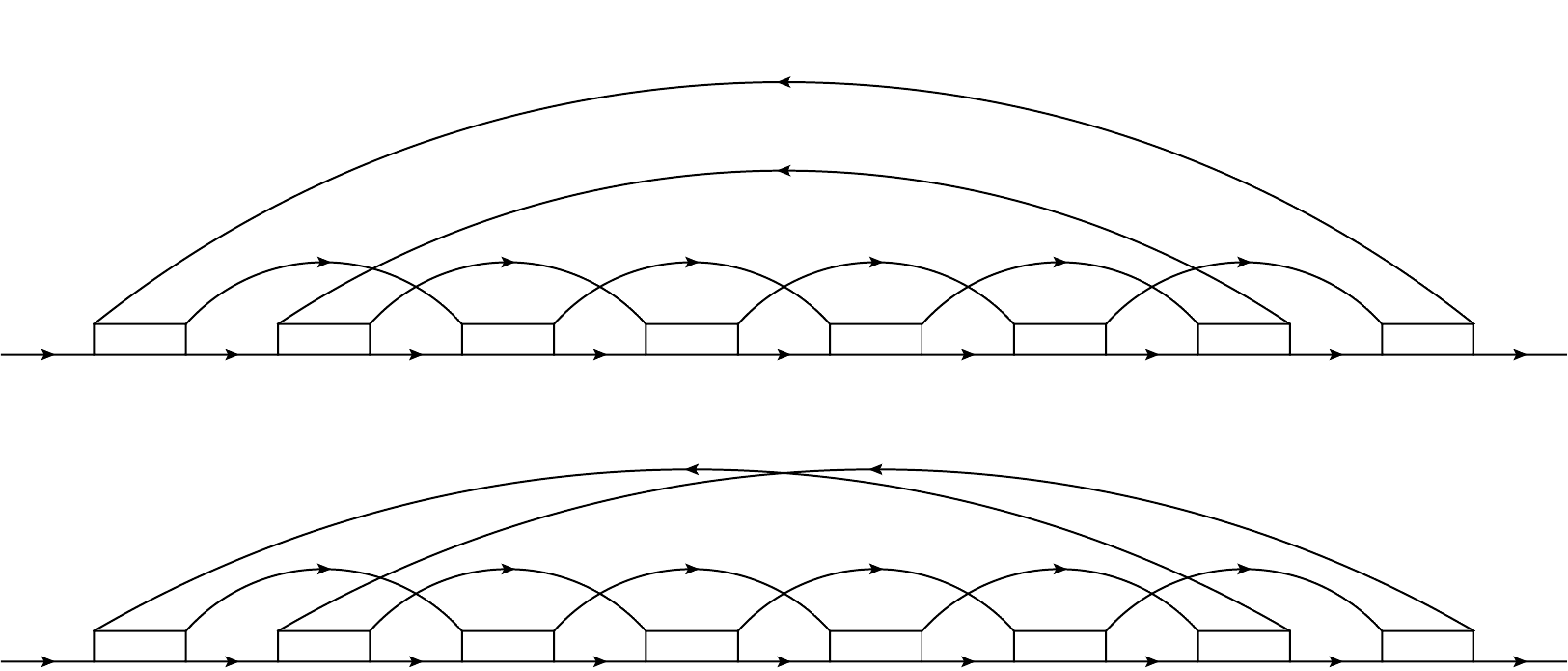}
\caption{Two diagrams contributing to $G_N$ obtained by closing the three-body T-matrix diagrams of Fig.~\ref{fig:T3diagrams} with two spin-up hole propagators in two different ways. 
}
\label{fig:dominant}
\end{figure}

\begin{figure}
\includegraphics[scale=0.38,width=0.99\columnwidth]{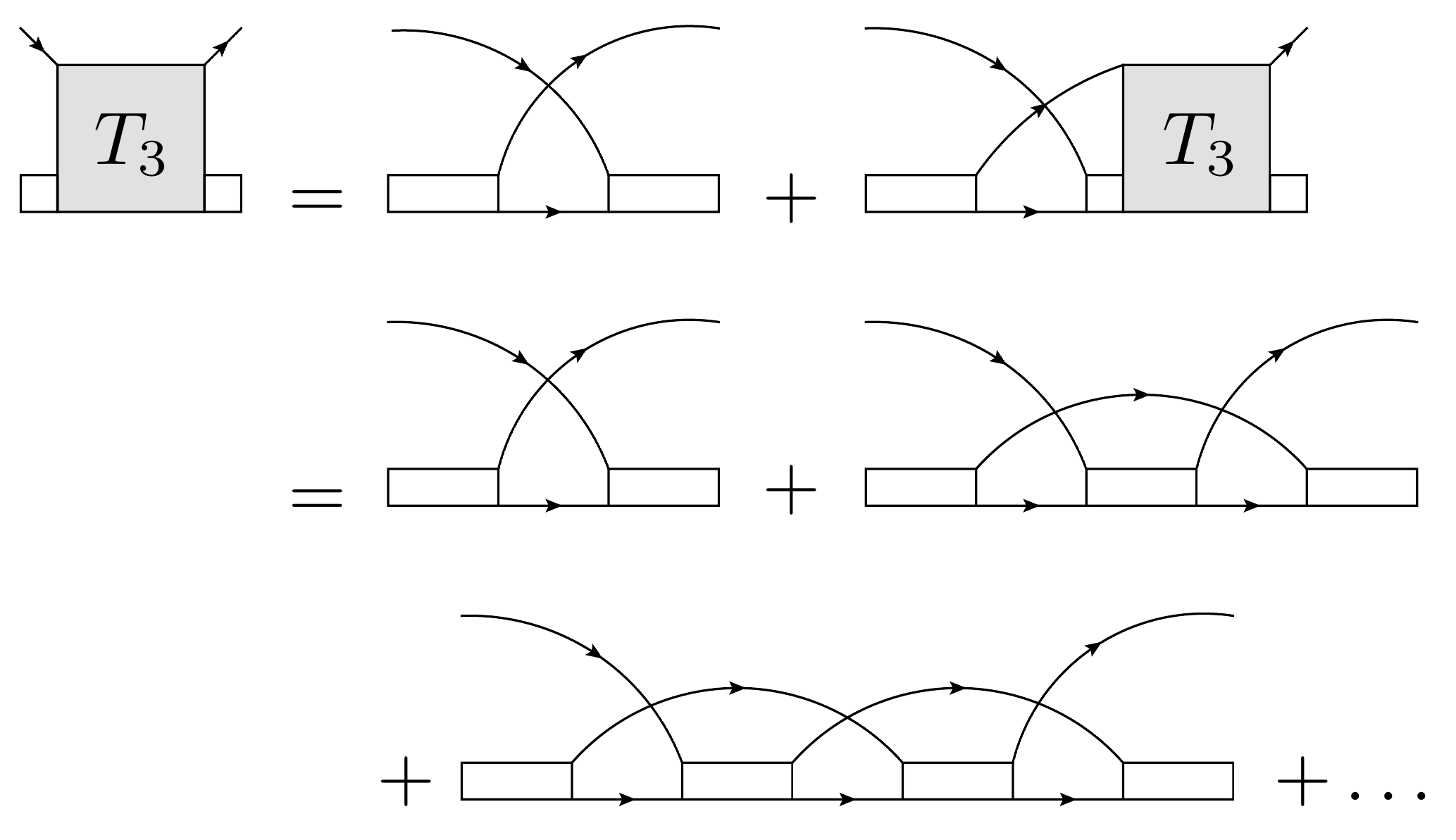}
\caption{The three-body T-matrix diagrams,
  describing the scattering between the impurity and two fermions.
}
\label{fig:T3diagrams}
\end{figure}

\begin{figure}
\includegraphics[scale=0.38,width=0.99\columnwidth]{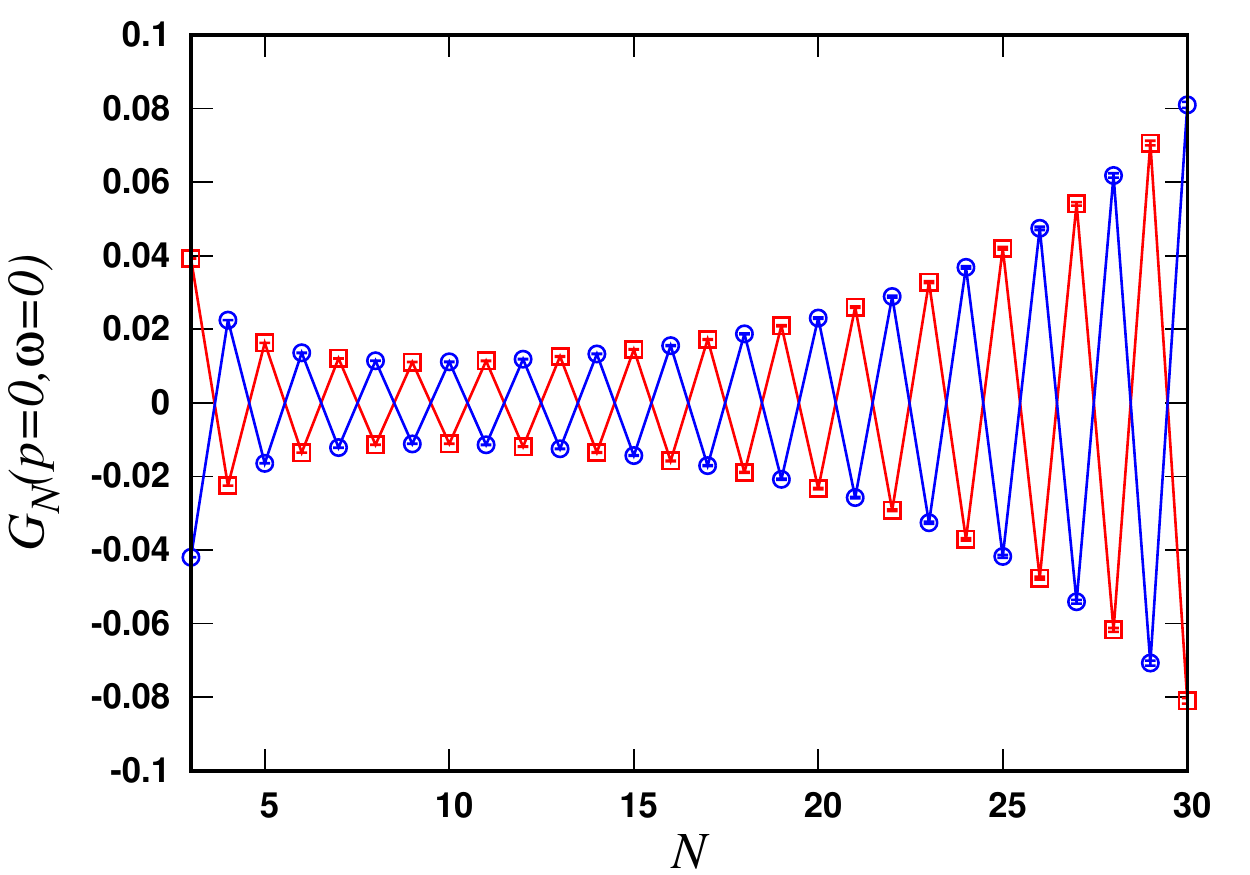}
\caption{The contribution  to $G_N(p=0,\omega=0)$ from the two diagrams shown in Fig.~\ref{fig:dominant}
as a function of the diagram order $N$.
The red squares and blue circles correspond to the contribution of the top and bottom diagram shown in Fig.~\ref{fig:dominant}, respectively. 
}
\label{fig:G_N_T3_omega0}
\end{figure}

\begin{figure}
\includegraphics[scale=0.38,width=0.99\columnwidth]{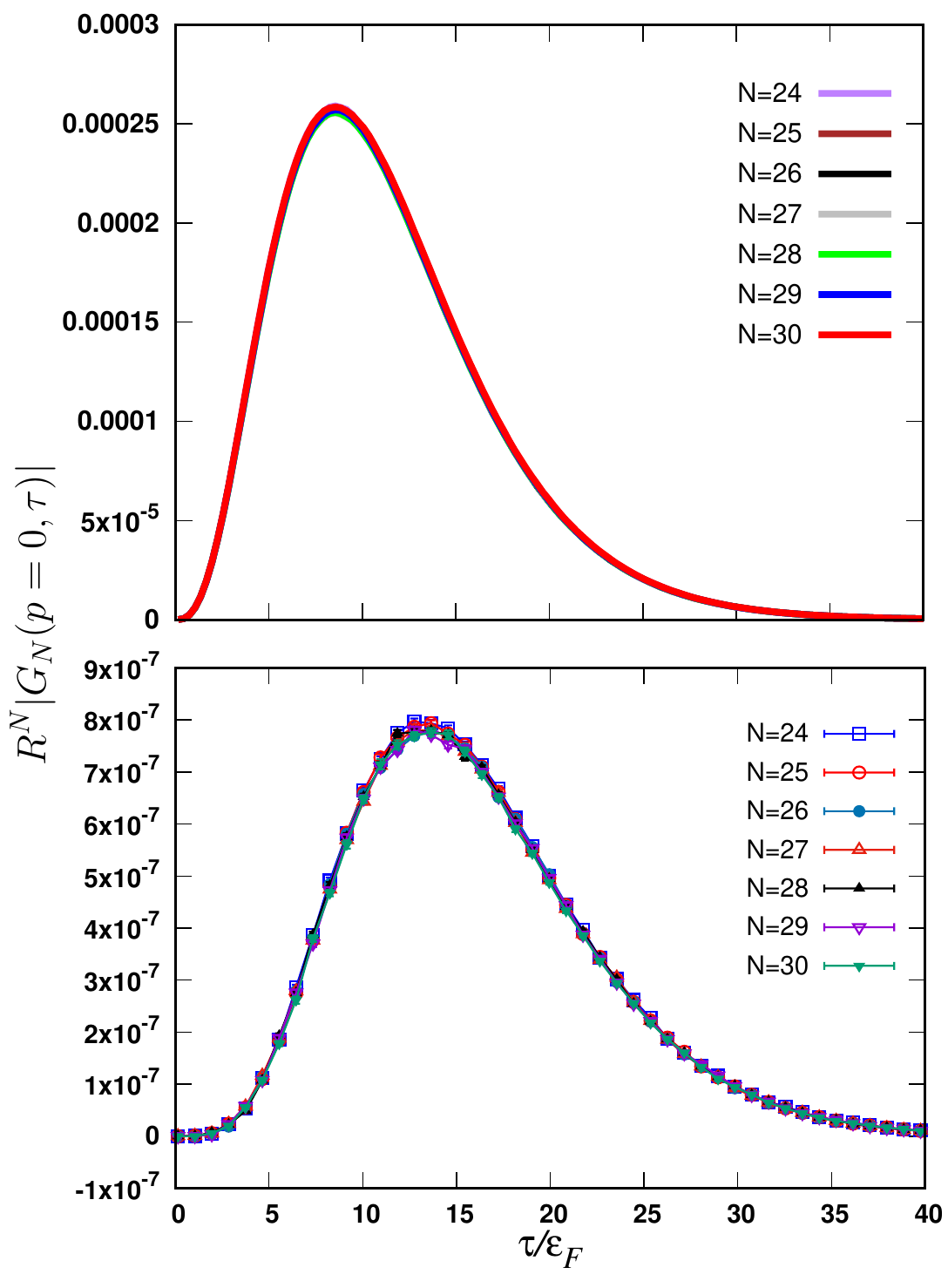}
\caption{The function $R^{N} |G_N(p=0,\tau)|$  for the first dominant diagram 
(upper panel)
and for the sum of the two dominant diagrams
(lower panel)
shown in Fig.~\ref{fig:dominant},
with $R = 0.8782(15)$. 
The width of the lines in the upper panel reflects the statistical error bars.
We again observe collapse of the data within the error bars, supporting the behavior Eq.~(\ref{eq:RNF}) at large enough $N$. 
}
\label{fig:G_N_T3_tau30}
\end{figure}

In order to understand how 
such an exponential divergence of the diagrammatic series can arise, we consider  the two diagrams shown in Fig.~\ref{fig:dominant}.
These diagrams have been considered before in the polaron problem~\cite{CGL2009,Vlietinck13}.
They can be viewed as three-body T-matrix diagrams, shown in Fig.~\ref{fig:T3diagrams}, closed with two hole propagators.
The three-body T-matrix describes the scattering between the impurity and two particles of the Fermi sea
(and also, in the strong-coupling limit, between a dimer and a fermion);
these are the only diagrams for the three-body problem in vacuum~\cite{Leyronas4corps}.

 Figure~\ref{fig:G_N_T3_omega0} shows the contribution to $G_N(p=0,\omega=0)$ of the two diagrams shown in Fig.~\ref{fig:dominant} as a function of $N$. 
Note that the contributions of these two diagrams nearly cancel each other, as noted before in Refs.~\cite{Combescot_Giraud,Vlietinck13}.
It turns out that those two diagrams follow the asymptotic behavior of Eq.~(\ref{eq:RNF}). 
When fitting the exponential increase in the range  $N=24-30$ for the first dominant diagram, we get $R = 0.8782(15)$.  This value is consistent with the value of $R$ obtained previously for the sum of all diagrams.  Figure~\ref{fig:G_N_T3_tau30} shows  the function $F(\tau) = R^{N} |G_N(p=0,\tau)|$ for the first dominant diagram 
and the sum of the two dominant diagrams.
 We clearly observe a collapse of the data within the statistical error bars. We conclude that at unitarity, the leading asymptotic behavior of the series, $(-R)^{-N}$, comes from the diagrams shown in Fig.~\ref{fig:dominant}.

Note that this conclusion 
is not related to the observations of Ref.~\onlinecite{Vlietinck13},
where
these diagrams (without the initial and final $G^{0}_\downarrow$ propagators) were identified as {\it qualitatively} dominant diagrams when sampling all possible self-energy contributions at {\it any} given order
 through DiagMC, 
their contribution being 
{\bl fairly} large compared to other diagrams 
of the same order because they contain a minimal number of backward propagating $G^{0}_\uparrow$ lines~\cite{Vlietinck13}.

\begin{figure}
\includegraphics[scale=0.38,width=0.99\columnwidth]{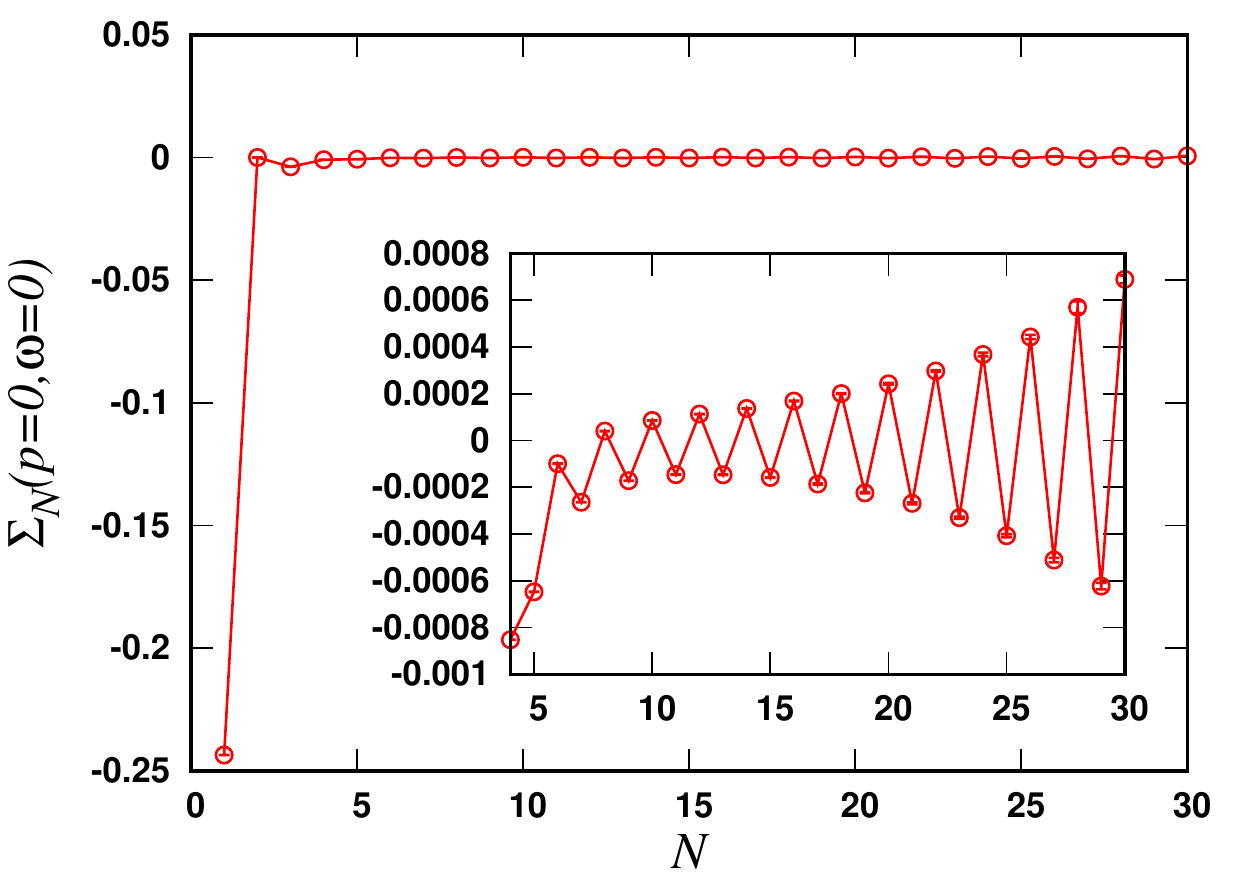}
\caption{The $N$-th order contribution $\Sigma_N$ to the self-energy at zero momentum and zero frequency for $k_F a = \infty$. Similar to $G_N$, the data shows an exponential and sign-alternating increase at high enough order. }
\label{fig:S_N_omega0}
\end{figure}

Calculation of  $\Sigma_N(p=0, \tau)$ and $\Sigma_N(p=0, \omega=0)$ reveal behavior very similar to the behavior of $G_N$. We again observe the behavior given in Eq.~(\ref{eq:RNF}) with the same value of $R$ as for the series for  $G$. Figure~\ref{fig:S_N_omega0} shows the contributions $\Sigma_N(p=0,\omega=0)$.

\begin{figure}
\includegraphics[scale=0.38,width=0.90\columnwidth]{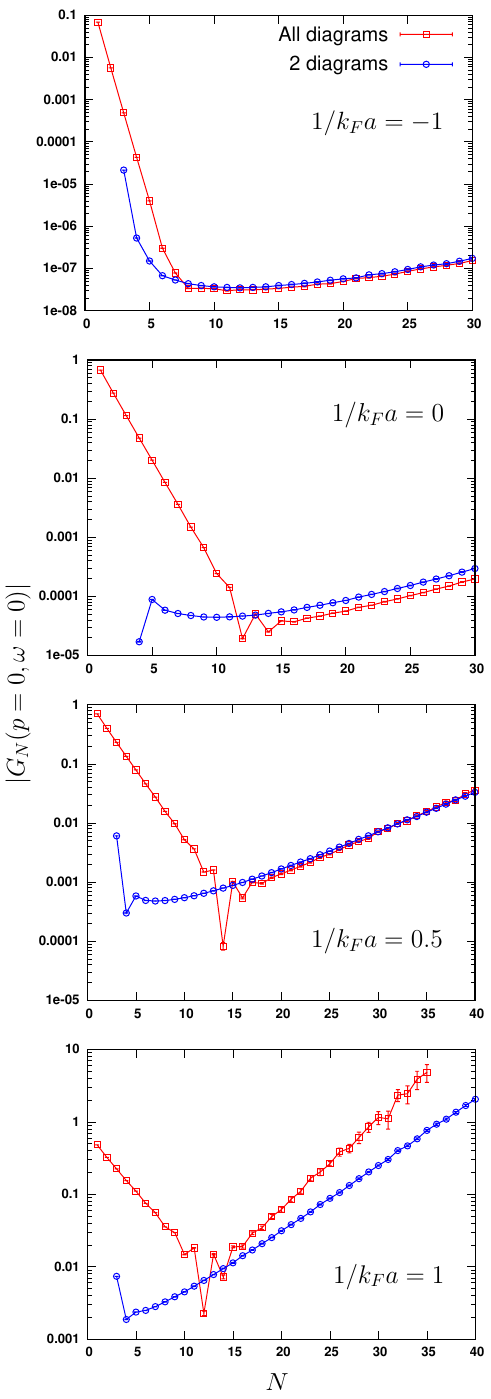}
\caption{The contribution $|G_N(p=0,\omega=0)|$ to the 
impurity propagator
as a function of the diagram order $N$ for the sum of all diagrams (red squares) and for the sum of the two diagrams (blue circles) shown in Fig.~\ref{fig:dominant}. The four panels correspond to different values of $1/(k_F a)$.  
}
\label{fig:G_N_omega0_a}
\end{figure}

While we focused on the unitary limit so far,
let us now look at the large-order behavior for different interaction strengths.
We find again an exponential divergence $G_N \sim (-R)^{-N}$ with $R<1$, but for strong coupling the value of $R$ is not determined any more by the three-body diagrams.
This is seen in Fig.~\ref{fig:G_N_omega0_a} which shows the contribution $|G_N(p=0,\omega=0)|$ as a function of $N$, on the one hand for all the diagrams, and on the other hand for the sum of the 2 diagrams shown in Fig.~\ref{fig:dominant}, for four values of $k_F a$.
In all cases, ${\rm log}\, |G_N|$ increases linearly with $N$,
but the slopes of the two curves clearly do not agree any more
for $1/(k_F a) = 1$, which is on the dimeron side
of the polaron to dimeron transition taking place at $1/(k_F a)_c = 0.87(2)$~\cite{Vlietinck13}.
In the other three cases, the slopes agree, although only marginally for $1/(k_F a) = 0.5$, where we pushed the calculation to order 40.
The values of $R$ are shown in Table~\ref{tab:R}~\footnote{We used a fitting range starting at $N_{\rm min}=24$, except in the cases [$1/(k_F a) = 0.5$] and [$R_{2 \,{\rm diag}}$, $1/(k_F a) = 1$] where we used $N_{\rm min}=30$.}.

\begin{table}
\begin{tabular}{|c|c|c|}
\hline
$1/(k_F a)$ & $R_{\rm all}$ & $R_{2\,{\rm diag}}$
\\
\hline
-1  &  \ \ \ 0.890(5)  \ \ \ &  \ \ \ 0.892(5)\ \ \ 
\\
\hline
0  &  0.878(3) &  0.879(3)
\\
\hline
0.5  &  0.850(4) &  0.857(2)
\\
\hline
1  &  0.752(11)  &  0.809(2)
\\
\hline
\end{tabular}
\caption{For various interaction strengths, comparison between
the convergence radius $R_{\rm all}$
from the sum of all diagrams,
and the convergence radius $R_{2\,{\rm diag}}$ from only the two~diagrams of Fig.~\ref{fig:dominant}.
\label{tab:R}}
\end{table}

We did not find classes of diagrams which 
would account
for this stronger divergence.
We considered the simplest diagrams contributing to the four-body propagators $T_4$ (closed in $3!$ possible ways) and to the five-body propagators $T_5$ (closed in $4!$ possible ways). They also show an exponential behavior as a function of $N$, but with a value of $R$ larger than 1, {\it i.e.}, they decrease exponentially with $N$.

{\bl
The exponential divergence found here for the polaron problem ({\it i.e.}, for a single $\downarrow$ particle)
is
entirely different from the large-order behavior
found in Ref.~\onlinecite{RossiEOS} for the many-body problem
({\it i.e.}, when both $\uparrow$ and $\downarrow$ particles have a finite density).
In the many-body case, the divergence is stronger, $\sim(N!)^{1/5}$,
so that the convergence radius is zero.
This is obtained from
a functional integral representation, with an action that depends not only on a fermionic field ---corresponding to the original fermionic particles--- but also on a bosonic field ---corresponding to pairs of fermion with opposite spin.
The factorial divergence essentially comes from 
contributions to the functional integral from the large bosonic field limit~\cite{RossiEOS,RossiThese}.
This is analogous to the Dyson collapse in QED~\cite{DysonCollapse,ParisiYukawa,ItzyksonParisiZuber,ItzyksonParisiZuber2,BogomolnyQed}.
In the polaron case, there is no representation in terms of a complex bosonic field. Physically, the number of bosonic $\uparrow\downarrow$ pairs cannot exceed one. Therefore, the mechanism responsible for the factorial divergence is absent in the polaron case.
}

\subsection{Power-counting argument} \label{subsec:power}

The exponential divergence with diagram order revealed by the above data contradicts a previous belief that the diagrammatic series should converge at fixed imaginary time~\cite{Goulko_Dark}.
This belief followed from the observation that time-ordering normally leads to a factorially convergent diagrammatic series: For fixed external imaginary time $\tau$, the contribution to $G_N(p,\tau)$
of any individual diagram is an integral  over {\it time-ordered} internal times, $\int_{0<\tau_1<\ldots<\tau_{2N}<\tau} f\,d\tau_1\ldots d\tau_{2N} $,
and 
{\it under the simplifying assumption that the integrand $f$ is bounded,}
this integral is bounded by $\tau^{2N}/(2N)!$ (omitting $N$-independent  prefactors);
since the number of order-$N$ diagrams is bounded by $N!$,
one conludes that $|G_N|$ is bounded by $(\tau/2)^{2N} / N!$. 
This naive conclusion is in contradiction with our numerical results.

We thus need to perform a more careful analysis, 
without making the above simplifying assumption.
We will see that in the present case of zero-range interactions in three-dimensional continuous space, the aforementioned effect of the time-ordering is exactly compensated by the effect of the short-time divergences of the propagators Eqs.~(\ref{eq:Gshort},\ref{eq:Gamshort}).

As a preliminary exercise, let us consider the simple integral
\be
I_n = \int _{0<\tau_1<\ldots<\tau_{n}<\tau} d\tau_1\ldots d\tau_{n}.
\ee
The integral can be evaluated exactly,
\be
I_n = \tau^n/n!
\label{eq:In}
\ee
Let us show how this behavior follows from a heuristic argument
 (before generalizing the argument to the polaron self-energy).
In the integral $I_n$, typically, the time-ordered variables $\tau_1,\ldots,\tau_n$ are spread in a roughly uniform way between 0 and $\tau$.
Therefore, each $\tau_i$ is effectively restricted to an interval of length $\sim \tau/n$. This leads to the estimate $I_n \sim (\tau/n)^n$, which agrees with the exact result Eq.~(\ref{eq:In}), up to a factor $({\rm constant})^n$ which is missed by this simple argument.

\begin{figure}
\includegraphics[scale=0.38,width=0.99\columnwidth]{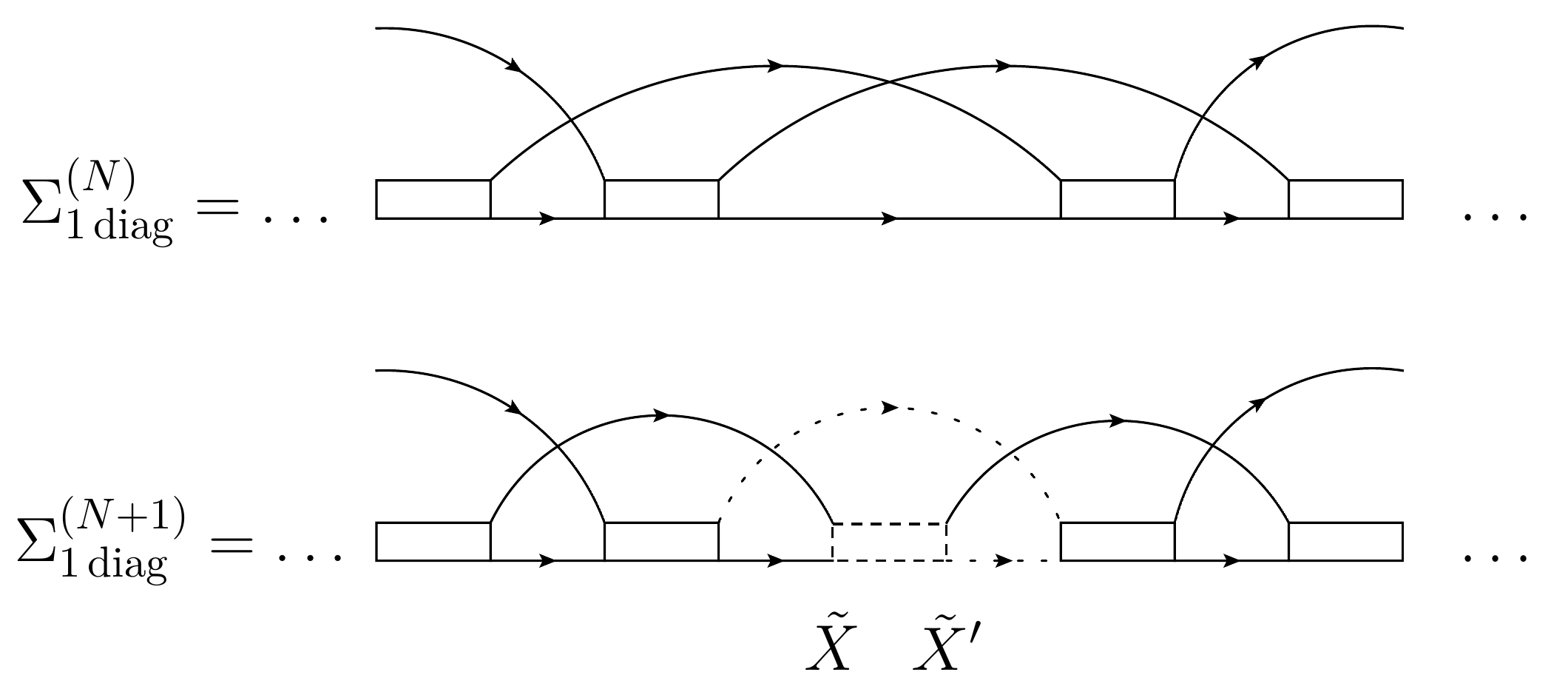}
\caption{For one of the three-body diagrams, increasing the order from $N$ to $N+1$ amounts to inserting
the building block represented in dotted lines.
\label{fig:power_count}}
\end{figure}

We now apply a similar kind of argument to the order-$N$ self-energy contribution $\Sigma^{(N)}_{\rm 1\,diag}$
of one of the two ``three-body diagrams''.
Once again,
let us start from the following 
assumption:
Typically, 
the time-ordered internal times $\tau_1\ldots\tau_{2N}$ are roughly uniformly spread between 0 and $\tau$,
so that
all the time-lengths $\tau_{\rm destination} - \tau_{\rm origin}$ of the lines in the diagram (either $G^0$ or $\Gamma^0$ lines)
are of the same order of magnitude $\Delta\tau(N)$.
Since the total time-length of the backbone $\sim 2 N\, \Delta\tau(N)$ has to match the external time $\tau$, we have $ \Delta\tau(N) \ll \tau$ for large $N$.
Now let us consider the ratio 
$\Sigma^{(N+1)}_{\rm 1\,diag} / \Sigma^{(N)}_{\rm 1\,diag}$.
We can view the order-$(N+1)$ diagram as the order-$N$ diagram with an additional structure
\begin{equation}
\includegraphics[width=0.3\columnwidth]{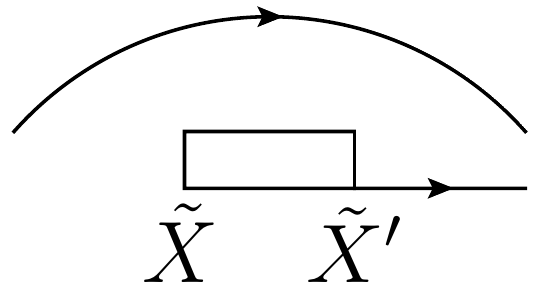} 
\end{equation}
inserted in the middle of the diagram,
see Fig.~\ref{fig:power_count}.
Accordingly, there are two additional internal space-time variables $\tilde{X}$ and $\tilde{X}'$.

From our assumption,
the integral over
$\tilde{X}$
is effectively restricted to a small volume (in space-time) $\Vr$ of order
$(\Delta\tau)^{5/2}$
[we will consider that $\Delta\tau(N)$ and $\Delta\tau(N+1)$ are close enough to neglect the $N$-dependence of $\Delta\tau$].
Indeed, the integral over $\tilde{\tau}$ is restricted to a small interval of length $\sim\Delta\tau$;
the integral over $\tilde{\mathbf{r}}$ is hence effectively resticted to a ball of radius $\sim\sqrt{\Delta\tau}$,
because the propagators decrease exponentially outside this ball
[see Eqs.~(\ref{eq:Gshort},\ref{eq:Gamshort})];
the volume of this ball in three dimensions is $\sim(\Delta\tau)^{3/2}$
which gives $\Vr \sim (\Delta\tau)^{1+3/2}$.
The same argument applies to the integral over $\tilde{X}'$,
which is effectively restricted to a volumes $\Vr'$, again of order
$(\Delta\tau)^{5/2}$.

This means that increasing the order has a cost: The two new internal variables have to fit into  small regions, which suppresses the result (as we have already seen for the preliminary exercise). Here the corresponding small multiplicative factor is $\Vr\,\Vr'\sim (\Delta\tau)^5$.
However, this is not the entire story:
Increasing the order by one also means 
adding three extra lines ---two $G^0$ lines and one $\Gamma^0$ lines (the dotted lines in Fig.~\ref{fig:power_count}).
These propagators have large values, of order
$1/(\Delta\tau)^{3/2}$ for the $G^0$ lines,
and $1/(\Delta\tau)^2$ for the $\Gamma^0$ line
[using again Eqs.~(\ref{eq:Gshort},\ref{eq:Gamshort}), where the exponentials are tpically $\sim 1$].
The resulting enhancement factor is $\sim 1/(\Delta\tau)^{2\times3/2 + 2}
= 1/(\Delta\tau)^5$,
exactly canceling out the above suppression factor coming from the smallness of the integration regions.
We conclude that $\Sigma^{(N+1)}_{\rm 1\,diag} / \Sigma^{(N)}_{\rm 1\,diag} \sim (\Delta\tau)^0 \sim 1$,
which suggests an exponential dependence of $\Sigma^{(N)}_{\rm 1\,diag}$ with $N$.
This  implies $G^{(N)}_{\rm 1\,diag} \propto (-R)^{-N}$,
since $G^{(N)}_{\rm 1\,diag}$ is just $\Sigma^{(N)}_{\rm 1\,diag}$ with an extra $G^0$ line attached at each end.

To summarize, 
when increasing
the diagram 
order by one as shown
in Fig.~\ref{fig:power_count}, 
a peculiar compensation takes place:
The smallness of the integration regions for the new time variables (which follows from the time ordering)
is exactly compensated by the large values of the new propagators;
therefore
 the order of magnitude of the diagram remains unchanged.
This scale-invariance property is specific to zero-range interactions in 
three-dimensional
continuous space,
 for which the propagators have the ultraviolet divergences Eqs.~(\ref{eq:Gshort},\ref{eq:Gamshort}).

We note that the reality is somewhat more complex than the above assumption
of roughly uniform spreading of the internal times, but we will argue that this should not change the conclusion.
First, the lines near the two ends of the diagram have no reason to have the same time-length than the lines in the ``bulk'' of the diagram;
however, these ``boundary effects'' should 
not affect the leading-order scaling coming from the ``bulk'' of the diagram.
Second, even in the ``bulk'', there are typically some lines with a time-length much larger than the other ones, {\it i.e.},
one does not have a single chain of short lines, but rather several bunches of short lines, separated by longer lines;
we observed this by looking at a few configurations visited by the Monte~Carlo process. This can be understood as an ``entropy-energy'' compromise:
The system decides to loose in ``energy'' by having some longer lines (with a smaller value of the propagators), but winning in ``entropy'' by increasing the effective accesible phase-space.
A quantitative study of this interesting effect is beyond the scope of this paper.
The above scaling argument can be expected to remain valid, since most of the lines remain short.

{\bl 
The fact that the exponential divergence of the diagrammatic series comes from ultraviolet behavior is further supported by the following observation.
Suppose that we evaluate diagrams in momentum-time representation,
and we {\it introduce a cutoff $p_c$} on all internal momenta.
Then one obtains the bound
\begin{eqnarray}
|G_N(p,\tau)|  & \leq &  \alpha(\tau)\  \frac{C^N ~p_c^{3N}  \tau^{\frac{3N}{2}}}{
\sqrt{ (N-1)!}}
\label{eq:with_cutoff}
\end{eqnarray}
for some $C$ and $\alpha(\tau)$. 
We thus conclude that the series would be 
convergent if there was a momentum cutoff.  
To derive Eq.~(\ref{eq:with_cutoff}), we replaced for simplicity
$\Gamma^0$ by the vacuum two-particle propagator at unitarity,
$\Gamma^v(p,\tau) =  - 
4 \sqrt{\pi/(m^3 \tau)}
\ e^{-(\frac{p^2}{4m}-\mu-\varepsilon_F)\tau}$,
which has the same large-momentum/short-time behavior as the full $\Gamma^0$ [cf. Eq.~(\ref{eq:Gamshort})].
 One thus has  $|\Gamma^v(p,\tau)|   \leq   A\,e^{B \tau}/\sqrt{\tau}$ with $A$ and $B$ some constants.  Moreover, the single-particle propagators are bounded, $|G^0_{\sigma}(p,\tau)| \leq 1$. 
Hence the $N$ integrals over independent internal momenta can be simply bounded by a factor $\propto p_c^{3N}$, and the remaining integrals over internal times can be done analytically, leading to Eq.~(\ref{eq:with_cutoff}).
}

Finally we note that some classes of diagrams do have a contribution which vanishes factorially at large $N$. For example, if we consider all reducible diagrams built from the lowest order self-energy $\Sigma_1$ [i.e., 
$G_N = (G^0_{\downarrow})^{N+1} (\Sigma_1 )^{N}$ in momentum-frequency representation], these diagrams will vanish as $1/(N!)^{3/2}$.

\subsection{Time dependence} 
\label{subsec:time-dep}

\begin{figure}
\includegraphics[scale=0.38,width=0.99\columnwidth]{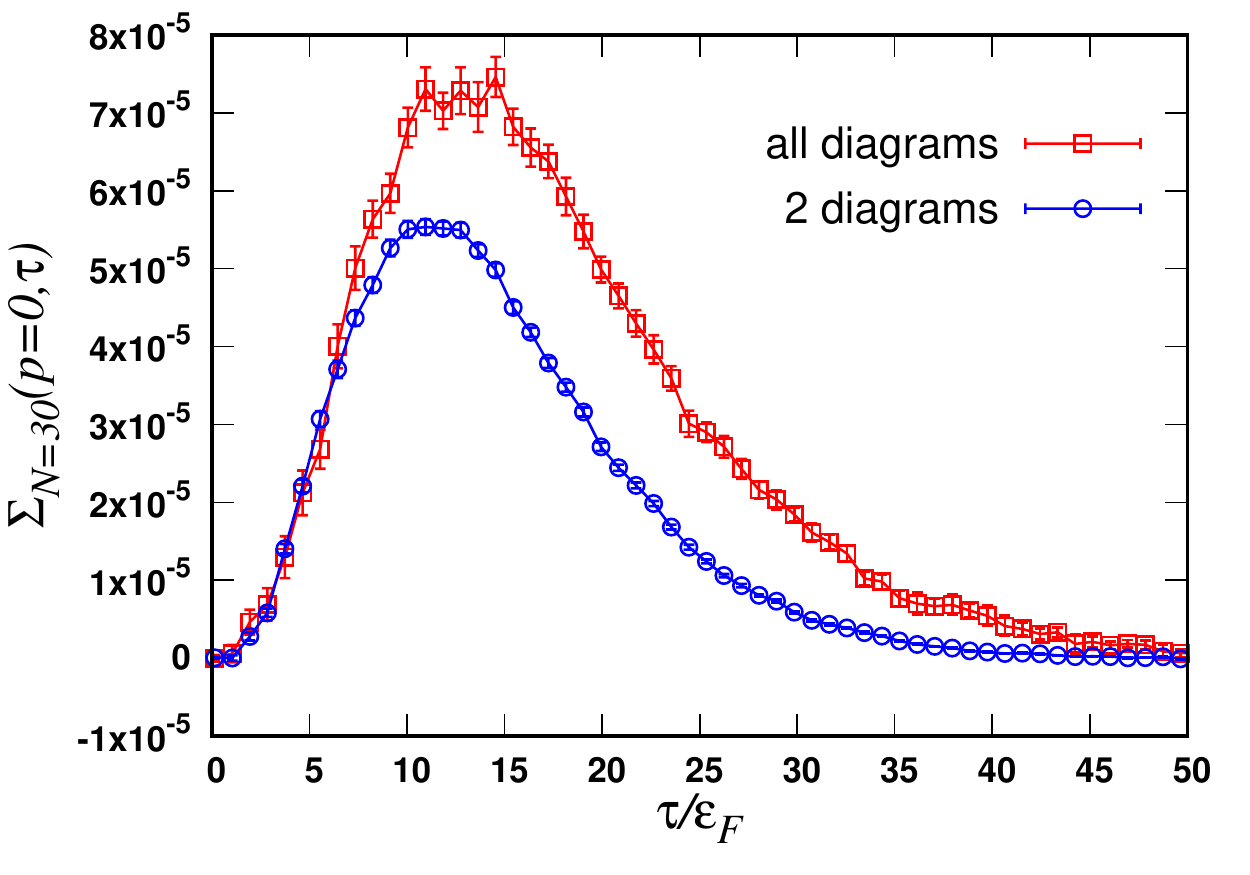}
\caption{The order-30 contribution to the self-energy at $p=0$ as a function of imaginary time $\tau$. Red squares: contribution of all diagrams, blue circles: contribution of only the self-energy insertions shown in Fig.~\ref{fig:dominant}.
}
\label{fig:Sigma_N30_tau50}
\end{figure}

The two diagrams of Fig.~\ref{fig:dominant} follow the asymptotic behavior of Eq.~(\ref{eq:RNF}), but they are not the only ones since they give a different function  $F_{{\rm 2\,diag}}(\tau) \neq F_{\rm all}(\tau) \equiv F(\tau)$.  To illustrate the difference, we show in Fig.~\ref{fig:Sigma_N30_tau50} the order $N=30$ contribution to $\Sigma(p=0,\tau)$ for all diagrams and for the two diagrams. 
We have investigated whether there exists a particular (simple) class of diagrams such that their sum reproduces the function $F_{\rm all}(\tau)$. 
 The conclusion of this search is that we could construct many different classes of  topologies  which lead to the same value of $R$, but we did not identify a simple class which reproduces the  function $F_{\rm all}(\tau)$. In the appendix, we give a number of examples of such topologies.    We leave the question whether there exists a simple class of diagrams which completely determines the asymptotic large-$N$ behavior as an open problem.

\section{Resummation}

Since the data clearly reveals a finite radius of convergence for the polaron diagrammatic series and since the physical answer is outside this radius, we can construct 
a conformal mapping in order to resum the series. 
Note that the Abelian resummation techniques used in Refs.[\onlinecite{Vlietinck13},\onlinecite{Goulko_Dark}] for the Fermi polaron problem can also deal with a finite radius of convergence and are an alternative way of resumming the series.
In contrast, the 
resummation methods used in Refs.~[\onlinecite{ProkofevSvistunovPolaronShort},\onlinecite{ProkofevSvistunovPolaronLong}] 
are strictly speaking not applicable given that the series diverges exponentially;
nevertheless, the results of Refs.~[\onlinecite{ProkofevSvistunovPolaronShort},\onlinecite{ProkofevSvistunovPolaronLong}] 
are consistent with the ones obtained here and in Ref.~\onlinecite{Vlietinck13},
which can be explained by the fact that the exponential divergence is rather weak (in the sense that the convergence radius is not much smaller than one)
and only develops at orders
$N\gtrsim 15$, which were not accessed in previous works.

We start by
interpreting the coefficients $\Sigma_N$
of the diagrammatic series for the self-energy $\Sigma$ as 
the Taylor coefficients of a function $\Sigma(z)$ of a formal parameter $z$:
\begin{equation}
\Sigma(z) := \sum_{N=1}^{\infty} \Sigma_N \,z^{N-1} \; ,
\label{eq:seriesz}
\end{equation}
where the physical self-energy corresponds to $\Sigma(z=1)$.
Given the asymptotic behavior
\begin{equation}
\Sigma_N \underset{N\to \infty}{\sim} (-1)^N R^{-N}  \; , 
\end{equation}
the series in Eq.~(\ref{eq:seriesz}) converges only for $|z|$ smaller than the radius of convergence $R$,
and the physical point $z=1$ is outside the convergence disk.
This can be cured
by
a conformal mapping, a method
used previously in the context of diagrammatic Monte~Carlo in 
Refs.~\cite{Profumo,RossiEOS,SimkovicCDet,Bertrand2}.
One  introduces a conformal mapping $z \mapsto w(z)$ such that $w_1=w(z=1)$ is inside the convergence disk of the transformed function $\tilde{\Sigma}(w) = \Sigma(z(w))$ in the $w$-plane.
Then, the physical result 
is obtained simply by evaluating the Taylor series of $\tilde{\Sigma}(w)$,
\begin{equation}
    \sum_{N=0}^{N_{\rm max}} \tilde{\Sigma}_N\, w^N,
\end{equation}
which converges in the limit $N_{\rm max}\to\infty$ at the physical point $w=w_1$.
{\bl Imposing $w(z=0)=0$ ensures that $\tilde{\Sigma}_N$ is a linear combination of $\Sigma_1,\ldots,\Sigma_{N+1}$.}

There are many different choices for the conformal mapping. 
We use
\begin{equation}
z(w) = \frac{A w}{(1-w)^{\alpha}}
\label{eq:mapping}
\end{equation}
with $\alpha > 0$.  This mapping is constructed such that the positive real axis in the $z$-plane is mapped onto the unit segment  in the $w$-plane, with 
$w(z=+\infty)=1$.
This guarantees that $0<w_1<1$,
and choosing $A = 2^{\alpha} R$ 
ensures that the singularity at $z=-R$ is mapped to
$w=-1$,
which is further away from the origin than $w_1$.
  The value of $\alpha$ can be chosen in the range $[0,2]$, fixing how much of the complex $z$-plane is mapped 
{\bl into the unit} disk. 
In what follows we take $\alpha=1$. We checked that the final result for the energy does not depend on $\alpha$ within the error bars.

\begin{figure}
\includegraphics[scale=0.38,width=0.99\columnwidth]{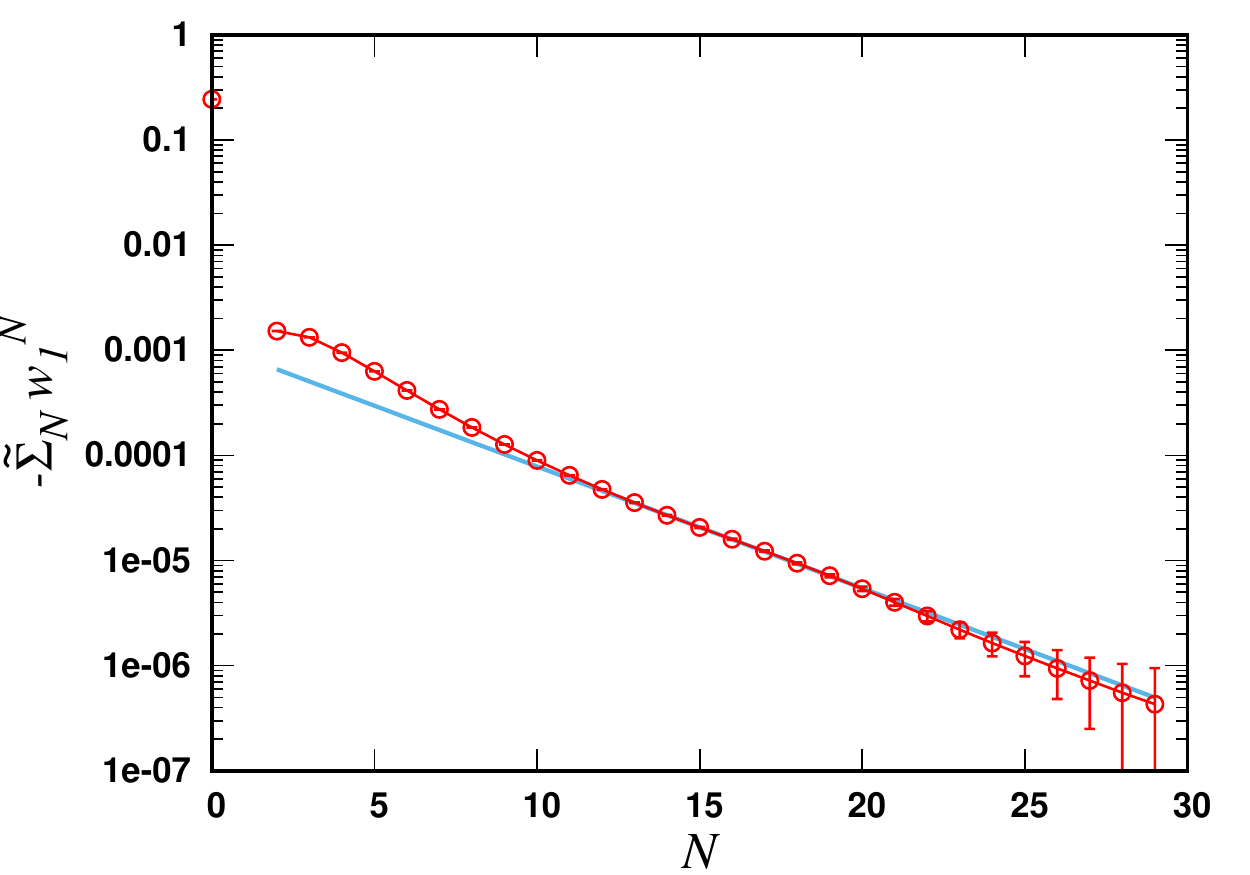}
\caption{Coefficients of the transformed series after conformal mapping.
We show  $-\tilde{\Sigma}_N(p=0,\omega=0) w_1^N$ as a function of order $N$.  The  line is an exponential fit. 
}
\label{fig:w3}
\end{figure}

After the conformal mapping, we 
observe that the series converges exponentially, see Fig.~\ref{fig:w3}.
This shows that all singularities were indeed mapped further away from the origin than $w_1$.
The fact that the series in $w$ is not sign-alternating indicates that 
the singularity $w_2$ nearest to the origin is
on the real positive axis. 
Fitting the tail gives $\tilde{\Sigma}_N \propto 1/w_2^N$ with $w_2 =  0.479(4)$, which is indeed larger than $w_1 = 0.3628$.
The corresponding singularity of $\Sigma(z)$ is at $z(w_2) = 1.61(3)$,
which we checked to be stable w.r.t. changing $\alpha$.

\begin{figure}
\includegraphics[scale=0.38,width=0.99\columnwidth]{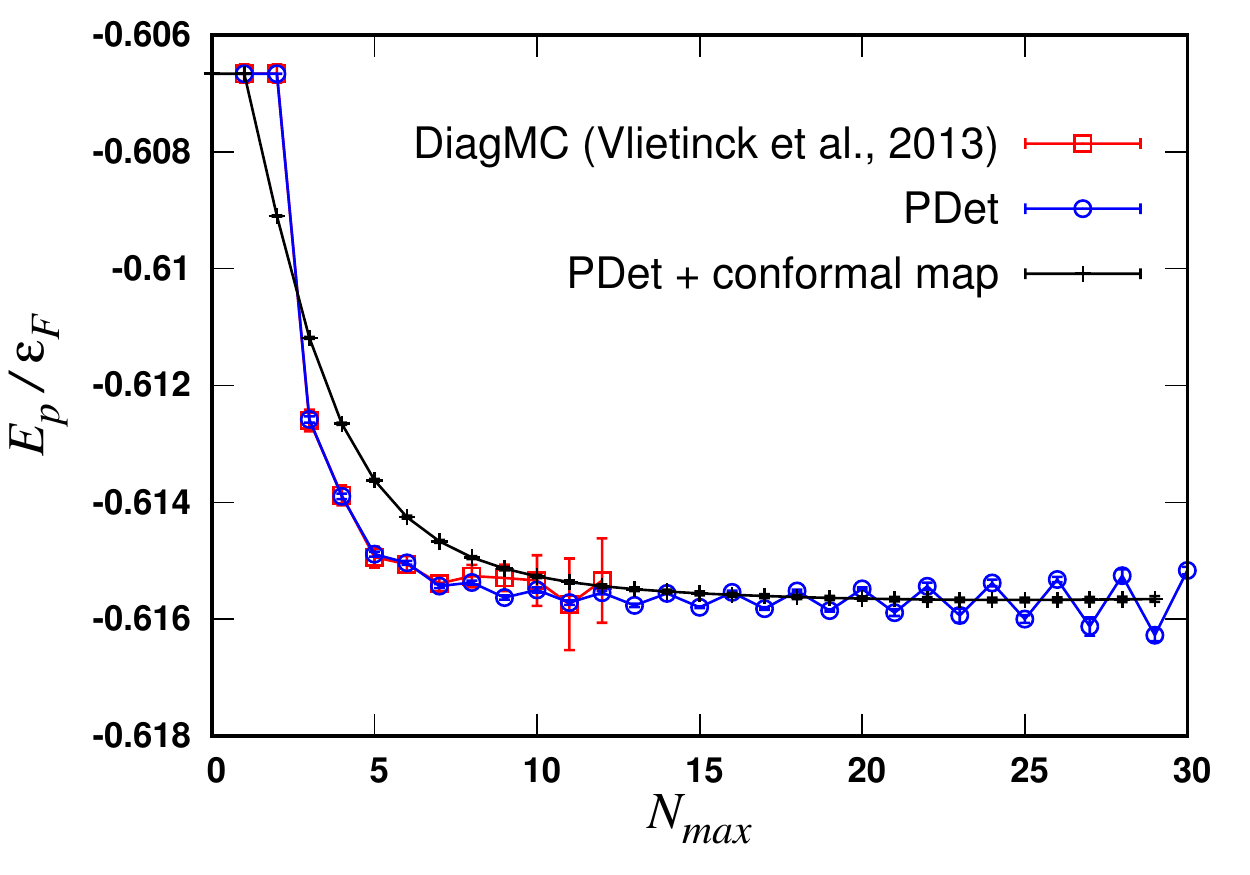}
\caption{The polaron energy $E_p$, 
as a function of the
maximal order $N_{\rm max}$
of the diagrammatic expansion of the self-energy.
While our results converge after conformal mapping (black crosses),
they diverge if this mapping is not applied (blue circles).
This divergence was not resolved in the earlier DiagMC results
from Ref.~\onlinecite{Vlietinck13}
(red squares). 
}
\label{fig:ener}
\end{figure}

The polaron energy $E_p$ is determined from the pole of the 
propagator $G$,
which gives
the implicit equation
in terms of the self-energy~\cite{ProkofevSvistunovPolaronShort}
\begin{equation}
E_p =  \Sigma(p=0, \omega = 0, \mu = E_p ).
\label{eq:ener}
\end{equation}
After applying the conformal mapping to the self-energy diagrammatic series, our results converge in the limit where the maximal diagram order $N_{\rm max}\to \infty$, see Fig.~\ref{fig:ener},
where we also show for comparison the results without the conformal mapping,
obtained with the new PDet algorithm, as well as with the older DiagMC method~\cite{Vlietinck13}.
We obtain
a polaron energy $E_{p} / \varepsilon_F = -0.61565(4)$.
This result is compared with earlier theoretical and experimental values in Table~\ref{tab:E}.
Our error bar is dominated by the systematic error from the Monte Carlo grid in imaginary time. 
The precision is strongly improved over 
the previous DiagMC results of Refs.~\cite{ProkofevSvistunovPolaronShort,ProkofevSvistunovPolaronLong,Vlietinck13}
and the hybrid path-integral/auxiliary-field quantum Monte Carlo result
of Ref.\cite{LeePolaron}.
The $\sim 1\%$ difference with the one particle-hole variational ansatz~\cite{chevy2006upa,Combescot2007} is much larger than our error bar,
and the agreement with the two particle-hole variational ansatz~\cite{Combescot_Giraud} is remarkable.
On the experimental side,  
we agree with the latest value obtained
 via radio frequency spectroscopy measurements of a strongly spin-balanced Fermi gas in a spatially uniform box potential~\cite{Boiling_MIT},
as well as with the earlier determinations of Refs.~\cite{ShinEOS,ZwierleinPolaron}.

\begin{table}
\begin{tabular}{|l|l|}
\hline
-0.61565(4) & this work
\\
\hline
-0.607 & one particle-hole variational ansatz~\cite{chevy2006upa,Combescot2007}
\\
\hline
-0.615(3) & diagrammatic Monte Carlo~\cite{ProkofevSvistunovPolaronShort,ProkofevSvistunovPolaronLong}
\\
\hline
-0.6156
& two particle-hole variational ansatz~\cite{Combescot_Giraud}
\\
\hline
-0.615(1) & diagrammatic Monte Carlo~\cite{Vlietinck13}
\\
\hline
-0.622(9) & 
lattice quantum Monte Carlo \cite{LeePolaron}
\\
\hline
-0.60(5) & experiment~\cite{Boiling_MIT}
\\
\hline
\end{tabular}
\caption{Polaron energy $E_{p} / \varepsilon_F$ at the unitary limit.
\label{tab:E}}
\end{table}

\begin{figure}
\includegraphics[scale=0.38,width=0.99\columnwidth]{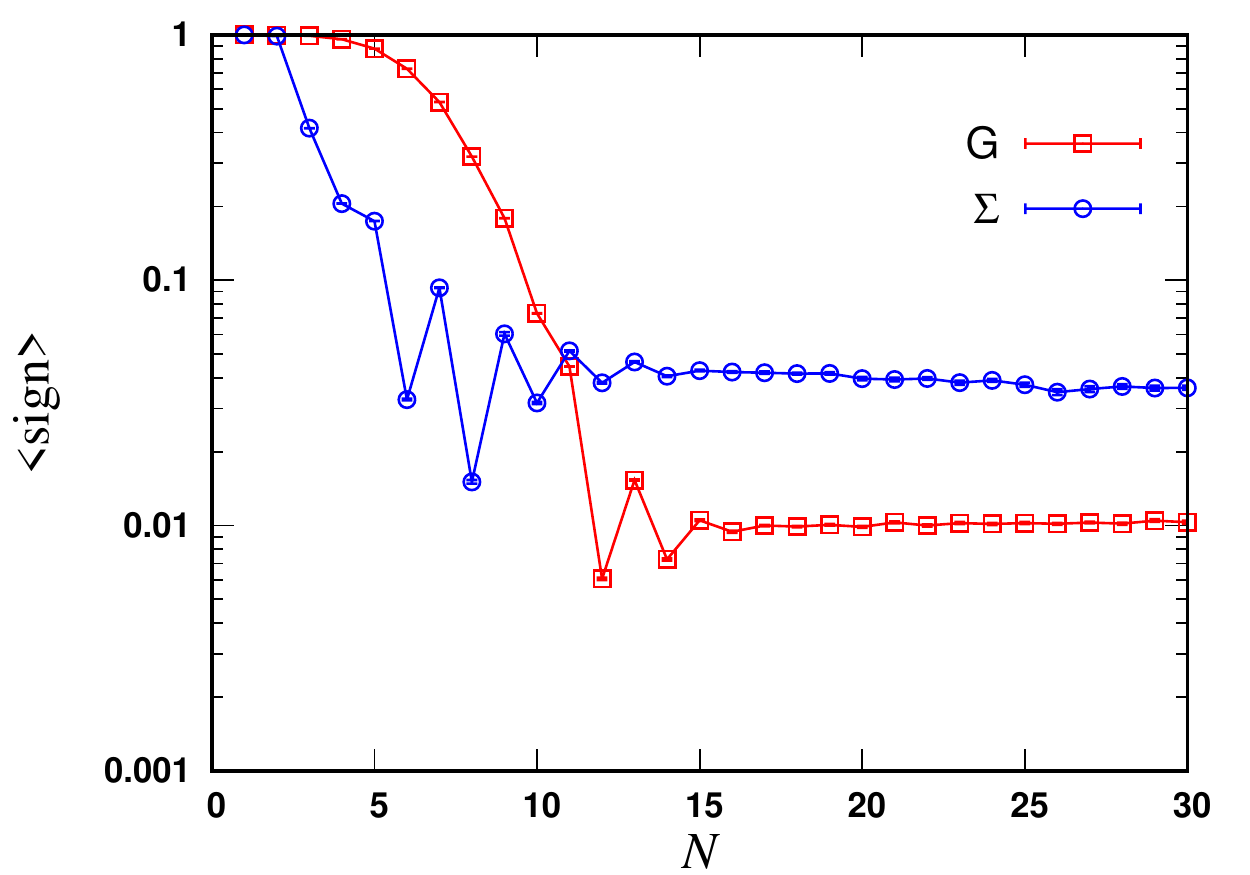}
\caption{Average sign as a function of order $N$ for $G(p=0,\omega=0)$ and $\Sigma(p=0,\omega=0)$. 
Remarkably, the average sign has a finite large-$N$ limit.
}
\label{fig:sign}
\end{figure}

\section{Efficiency of the algorithm}
We end with a quantitative discussion of the new algorithm's efficiency. Consider
the computation of a quantity $Q$, of diagrammatic expansion
$$\sum_{N=1}^\infty a_N.$$
The most relevant case for PDet is the self-energy $Q=\Sigma$, for fixed external variables, say $(p=0,\omega=0)$ for simplicity,
so that $a_N = \Sigma_N(p=0,\omega=0)$.

Denoting the set of space-time variables by $V_N$,
we have
\be
a_N = \int dV_N\ W(V_N);
\label{eq:aN_int_dV}
\ee
in the considered case of the self-energy with PDet, we have $V_N \equiv \{X'_1,X_2,X'_2,\ldots,X_N,X\}$
  and $W(V_N) = \tilde{B}(V_N) \tilde{S}(V_N)$, see Eqs.~(\ref{eq:Sigma_X},\ref{eq:Btilde},\ref{eq:Stilde}).
Equation~(\ref{eq:aN_int_dV}) can be rewritten as
\be
a_N = \la {\rm sign} \ra_N \ z_N
\ee
where $\la {\rm sign} \ra_N$ is the average sign corresponding to the Monte~Carlo process at order $N$,
$ \la {\rm sign} \ra_N = \la {\rm sign} \,W(V_N)\ra$
where the average is taken w.r.t. the Monte~Carlo weight $|W(V_N)|$,
and
\be
z_N = \int dV_N\ |W(V_N)|
\ee
is the total weight of the order-$N$ configuration space.

\subsection{Average sign}

A major aspect determining the efficiency of any
Monte Carlo algorithm for fermions is the behavior of the
average sign. A small average sign means that positive
and negative contributions nearly cancel out on average,
which amplifies the relative statistical error. In previous diagrammatic Monte Carlo algorithms for fermionic
many-body or polaron problems, the average sign tends
to zero in the large-order limit. This ``sign problem''
poses a fundamental limitation on the order that can be
reached within a given computational time. In contrast,
for PDet the average sign tends to a finite limit at large
order, as we see in Fig.~\ref{fig:sign}\,: Remarkably, the fermionic
sign does not cause any fundamental difficulty here.

While this observation is surprising at first, we can
understand it from the power-counting argument of
Sec.~\ref{subsec:power}. That argument suggested that $|a_N| \sim (1/R)^N$
at large $N$ , with $R$ determined by any of the two three-body diagrams. The same argument gives $z_N \sim (1/R)^N$
with the same $R$: Indeed, it does not matter that we
consider the integral of $|W|$ instead of the integral of
$W$, because the short-time expressions of the propagators Eqs.~(\ref{eq:Gshort},\ref{eq:Gamshort}) are sign-definite. This explains that
$\la{\rm sign}\ra_N = a_N /z_N$ has a finite limit for $N\to\infty$.

\subsection{Computational complexity}
A natural way of summarizing all aspects of computational complexity for a numerical algorithm is to determine how the computational time $t$ scales with the error~$\epsilon$. Here
$\epsilon$ is the difference between the computed value and the
exact result, coming from both statistical and systematic
errors. As we will see, the scaling is only polynomial
{\rsub in $1/\epsilon$}
for PDet,
\be
t = O\!\left(  1/\epsilon^\nu \right).
\label{eq:t_vs_eps}
\ee

Such a scaling was derived in Ref.~\cite{RossiComplexity}
for the CDet algorithm,
and
we can reuse most of the analysis presented there,
modulo the following three differences,
which will change the expression of the exponent $\nu$:
\vskip 0.15cm
\noindent {\it (i)}
the number of operations per Monte~Carlo step increases only polynomially with $N$ for PDet, instead of the $3^N$ scaling of CDet
{\rsub
(because for the polaron problem, disconnected diagrams do not exist,
so that the CDet recursive procedure to eliminate disconnected diagrams is not required for PDet)};
\\{\it (ii)} 
while the average sign decreases exponentially with
$N$ for a full many-body problem with CDet, it has a
finite large-$N$ limit for the Fermi-polaron problem with
PDet;
\\{\it (iii)}
while Ref.~\cite{RossiComplexity} restricted for simplicity to the case of a convergent series ({\it i.e.} to
small enough interaction
for the Hubbard model), here we have to consider the case of a divergent series, resummed by conformal mapping.
\vskip 0.15cm
\noindent 
It 
was already stated in Refs.~\cite{SimkovicCDet,Bertrand2}
that point {\it (iii)} does not invalidate the scaling~(\ref{eq:t_vs_eps});
here we will justify this in some detail and
show how this modifies the exponent $\nu$ (both for PDet and CDet).

Whereas the original series $\sum a_N$ diverges exponentially, $|a_N| \sim 1/R^N$ with $R<1$,
after conformal mapping one obtains a convergent series
\be
Q = \sum_{N=1}^\infty \tilde{a}_N
\ee
[in our case $\tilde{a}_N = \tilde{\Sigma}_{N-1}(p=0, \omega=0)\,w_1^{\phantom{1}N-1}$]
and $Q$ is computed by evaluating the truncated series
\be
Q^{(N_{\rm max})} = \sum_{N=1}^{N_{\rm max}} \tilde{a}_N
\ee
for some maximal order $N_{\rm max}$.
The transformed series converges exponentially,
\be
\tilde{a}_N = O(1/\tilde{R}^N)
\ee
with $\tilde{R}>1$.
Hence the truncation error is $\sim 1/\tilde{R}^{\Nmax}$.

To evaluate the statistical error, we have to return to the original coefficients $a_N$, which are the ones evaluated by Monte~Carlo.
Since the $\tilde{a}_N$ are linear combinations of the $a_N$,
we have
\be
Q^{(N_{\rm max})} = \sum_{N=1}^{N_{\rm max}} a_N\ F^{(\Nmax)}_N
\ee
Here the cofficients $F^{(\Nmax)}_N$ depend on the conformal map;
they necessarily tend to 1
for $\Nmax\to\infty$ at fixed $N$,
and they
typically smoothly decrease from nearly 1 to nearly 0
as a function of $N$ at fixed large $\Nmax$.
Neglecting correlations between the $a_N$,
the statistical error on $Q^{(N_{\rm max})}$ is given by
\be
\epsilon_{\rm stat}^2 \simeq \sum _{N=1}^{N_{\rm max}} \epsilon_{\rm stat}(N)^2\ \left(F^{(\Nmax)}_N \right)^2
\ee
with $\epsilon_{\rm stat}(N)$ the statistical error on $a_N$.
Since the $F^{(\Nmax)}_N$ are bounded (they are typically between 0 and 1),
we can simply use the bound
\be
\epsilon_{\rm stat}^2 \leq  \bar{\epsilon}_{\rm stat}^{\ 2} := \sum _{N=1}^{N_{\rm max}} \epsilon_{\rm stat}(N)^2.
\ee
The rest of the discussion is similar to Ref.~\cite{RossiComplexity}.
Given that $\la {\rm sign} \ra_N$ has a finite large-$N$ limit,
one finds that
$\bar{\epsilon}_{\rm stat} \sim (1/R)^{\Nmax} / \sqrt{t}$.
This scaling is related to the fact that
when resumming the divergent series $\sum a_N$,
there is necessarily a near-compensation between the contributions of different
$a_N$ to the resummed result,
so that the required relative accuracy on $a_N$ increases with $N$.
Choosing $\Nmax$ such that the truncation error is of the same order than $\bar{\epsilon}_{\rm stat}$
then leads to the result Eq.~(\ref{eq:t_vs_eps}) with the exponent
\be
\nu = 2 + 2\ \frac{{\rm log} (1/R)}{{\rm log}\,\tilde{R}}.
\label{eq:nu_pdet}
\ee
At the unitary limit, we have $\tilde{R} = w_2/w_1 \simeq 1.32$ and $R\simeq 0.88$, which gives $\nu \simeq 2.9$, a remarkably small value
(the best possible scaling for any Monte~Carlo computation being $\nu=2$).

For CDet, the result obtained in Ref.~\cite{RossiComplexity}
for the convergent-series case ($R>1$) is
$\nu = 2 + 2\,{\rm log} (3/R_C^{\phantom{C}2}) / {\rm log}\,R$,
where $R_C$ is such that $z_N \sim 1/R_C^N$ at large $N$
(discarding here the exotic case $R_C>\sqrt{3}$);
in the divergent-series case ($R<1$) the above discussion shows that we only need to replace $R$ with $\tilde{R}$ in the expression of the truncation error, which gives
\be
\nu_{\rm CDet} = 2 + \frac{{\rm log} (3/R_C^{\phantom{C}2})}{{\rm log}\,\tilde{R}}.
\label{eq:nu_cdet_resum}
\ee
From this expression, the PDet result Eq.~(\ref{eq:nu_pdet}) can be retrieved 
by removing the factor 3 and setting
$R_C=R$;
this follows 
from the above points {\it (i)} and {\it (ii)} respectively.

\section{Conclusion}

We introduced
 an 
algorithm
to
solve numerically the Fermi polaron problem with high precision.
With respect to the existing diagrammatic Monte Carlo algorithm~\cite{ProkofevSvistunovPolaronShort},
the progress is  substantial, both in terms of efficiency and of algorithmic simplicity.
The obtained high-order data have clarified a conceptual aspect of fundamental importance,
the large-order behavior of the diagrammatic series,
which is found to diverge exponentially at a rate determined by a single diagram. This peculiar situation is made possible by a compensation between the effects of time-ordering and of ultraviolet divergencies for the zero-range interaction in three dimensions.
This compensation also implies
that the average sign remains finite in the large-order limit,
which means that the fermionic sign does not cause any essential problem preventing to reach high orders.
The knowledge of the large-order behavior allows to resum the series
in an efficient and controlled way
by means of a conformal map,
as demonstrated by first illustrative results.

{\it Acknowledgements.}
We thank X.~Leyronas and N.~Prokof'ev for discussions. 
This work was granted access to the HPC resources of {\it MesoPSL} financed
by the Region Ile de France and the project {\it Equip@Meso} (reference
ANR-10-EQPX-29-01) of the programme Investissements d'Avenir supervised
by the Agence Nationale pour la Recherche.
F.W. acknowledges support from ERC grant {\it Critisup2} (Adv 743159).

\appendix*
    \section{Some diagrams contributing to the  large-order behavior} \label{app:diagrams}
    
\begin{figure}
\includegraphics[scale=0.38,width=0.99\columnwidth]{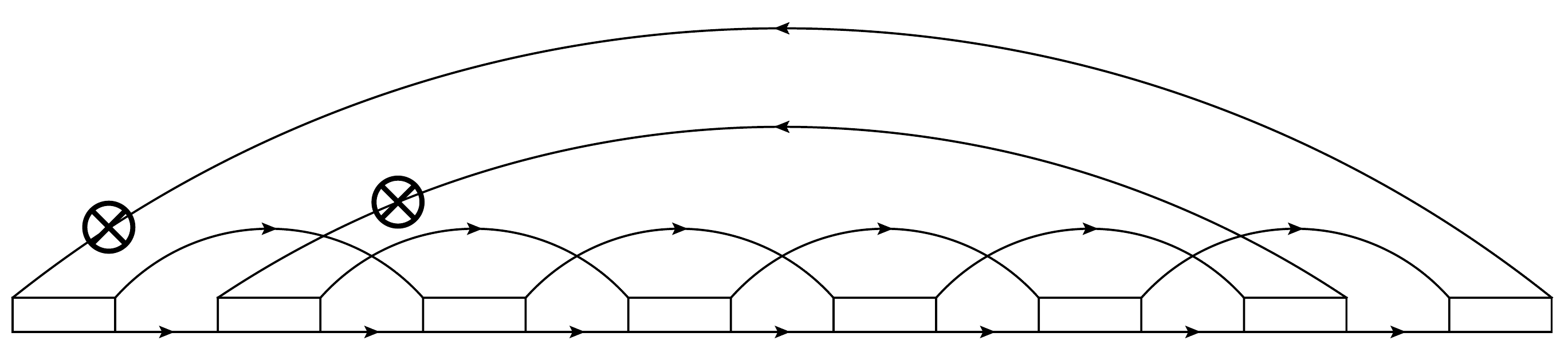}
\caption{Two self-energy diagrams contributing to the asymptotic large-order behavior given in Eq.~(\ref{eq:RNF}).
One topology is drawn explicitly. The second topology is obtained by interchanging the ends of the $G^0_{\uparrow}$-propagators marked with the symbol $\otimes$.
 The same two self-energy contributions are shown explicitly in Fig.~\ref{fig:dominant} (with two additional external $G^0_{\downarrow}$-lines). 
}
\label{fig:dominant_2}
\end{figure}

    We have numerically found that  the $N$-th order contribution $G_N(p=0,\tau)$ to the 
propagator,  or  $\Sigma_N(p=0,\tau)$ to the self-energy,  follows the asymptotic behavior given in  Eq.~(\ref{eq:RNF}) at large enough order. 
We investigated whether there exists a particular (simple) class of diagrams that is responsable for such a remarkable asymptotic behavior. As explained in the main text, the two diagrams shown in Fig. ~\ref{fig:dominant_2} already have  the same asymptotic behavior with the same value of $R$ but with a different function $F_{{\rm 2\,diag}}(\tau) \neq F_{\rm all}(\tau) \equiv F(\tau)$. We therefore considered additional classes of diagrams in the hope that their sum does not only give the same value of $R$ but also the same function $F_{\rm class}(\tau) = F_{\rm all}(\tau)$.  
    The conclusion of this search is that we could construct many different topologies  which lead to the same value of $R$, but we failed to identify a simple class which reproduces the  function $F_{\rm all}(\tau)$. In this appendix, we give a number of examples of such topologies. 
    
\begin{figure}
\includegraphics[scale=0.38,width=0.99\columnwidth]{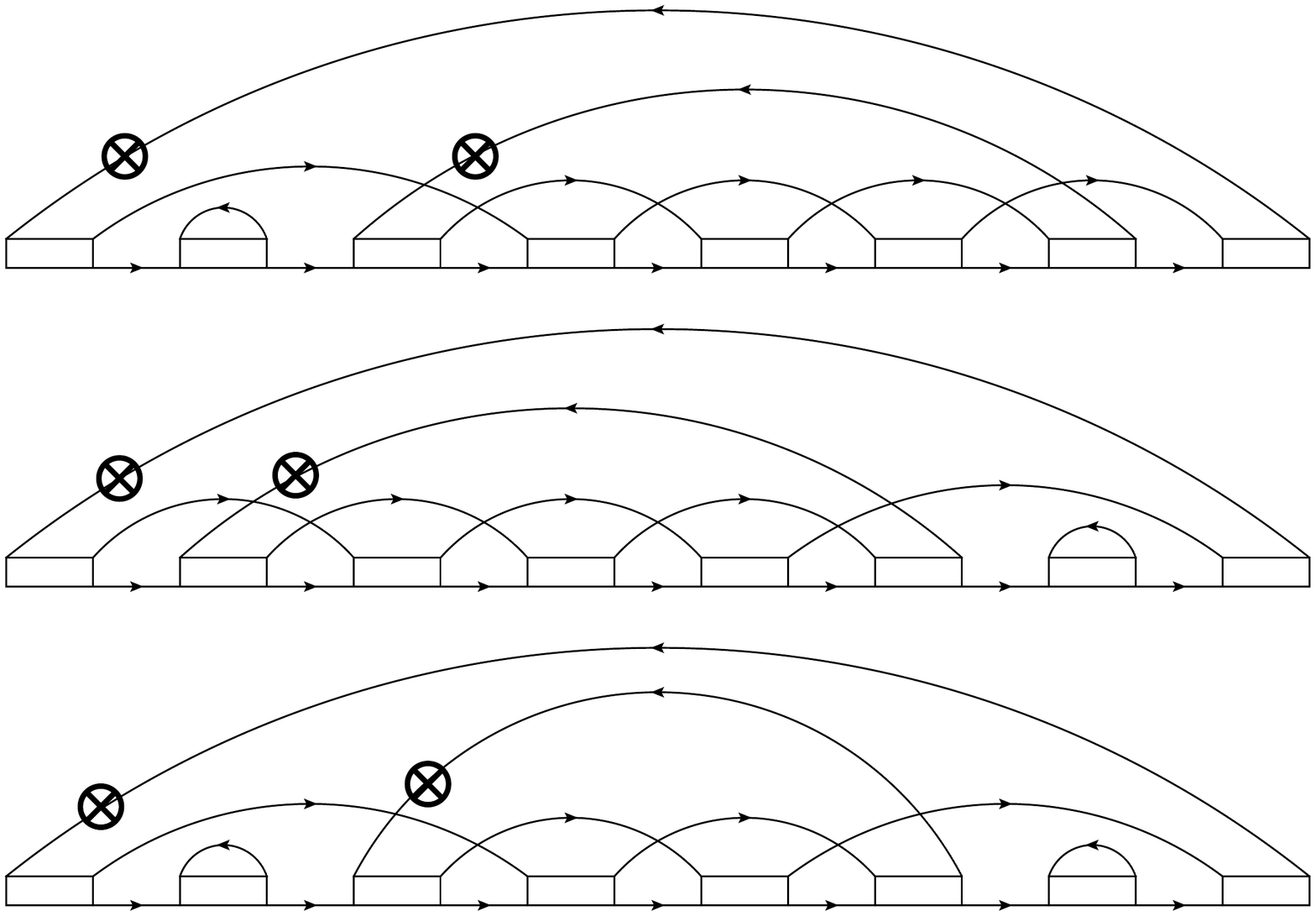}
\caption{A class of self-energy diagrams contributing to the asymptotic large-order behavior given in Eq.~(\ref{eq:RNF}).  Each diagram represents two topologies, following the rules explained in the caption of Fig.~\ref{fig:dominant_2}.
}
\label{fig:dressedbb1}
\end{figure}

\begin{figure}
\vskip 0.1cm
\includegraphics[scale=0.38,width=0.99\columnwidth]{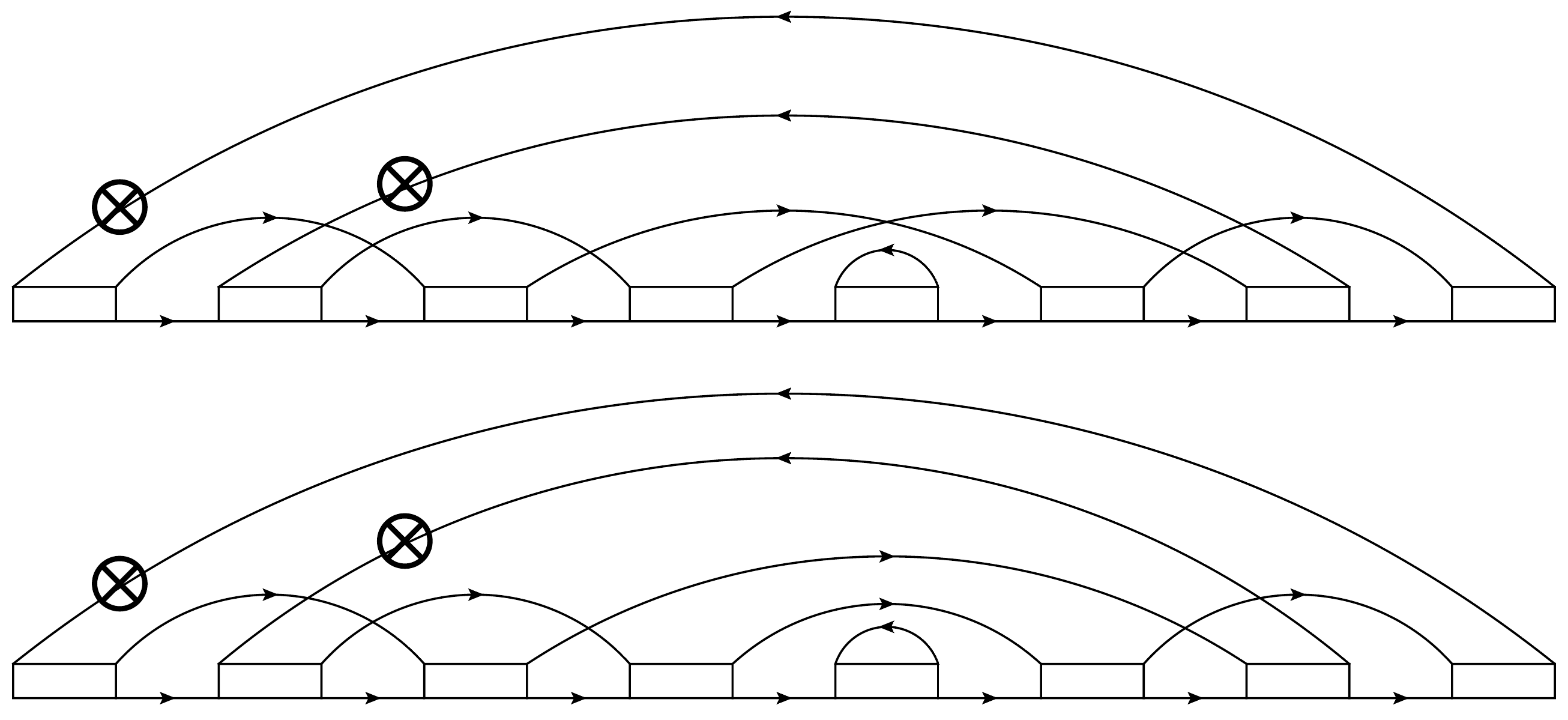}
\caption{Some more self-energy diagrams contributing to the asymptotic large-order behavior given in Eq.~(\ref{eq:RNF}).   These diagrams are similar to those of Fig.~\ref{fig:dominant_2}.
The difference is a self-energy insertion in the middle of the backbone, which effectively dresses one of the single-particle lines of the backbone (upper diagram) or one of the two-particle lines of the backbone (lower diagram). 
}
\label{fig:dressedbb2}
\end{figure}

A first class of diagrams we consider is shown in Figs.~\ref{fig:dressedbb1} and~\ref{fig:dressedbb2} .  
The diagrams shown in Fig.~\ref{fig:dressedbb1} are identical to the ones shown in Fig.~\ref{fig:dominant_2}, with the addition that some backbone lines are dressed. 
More specifically, the first and/or last $G^0_{\downarrow}$ of the backbone is dressed with a first order self-energy contribution $\Sigma^{(1)}$ [i.e., $\Sigma^{(1)}(r, \tau) = \Gamma^0(r,\tau) G^0_{\uparrow}(r,-\tau)$].
For the diagrams of Fig.~\ref{fig:dressedbb2}, one such $\Sigma^{(1)}$ contribution appears in the middle of the backbone, such that one single-particle propagator (top diagram in Fig.~\ref{fig:dressedbb2}) or one two-particle propagator (lower diagram in \ref{fig:dressedbb2}) appears to be partially dressed.  
All diagrams of Figs.~\ref{fig:dressedbb1} and~\ref{fig:dressedbb2} contribute to $F_{{\rm all}}(\tau)$.  
The contribution of the diagrams of Fig.~\ref{fig:dressedbb2}, however, is three orders of magnitude smaller than the contribution of those of Fig.~\ref{fig:dressedbb1}.

\begin{figure}
\includegraphics[scale=0.38,width=0.99\columnwidth]{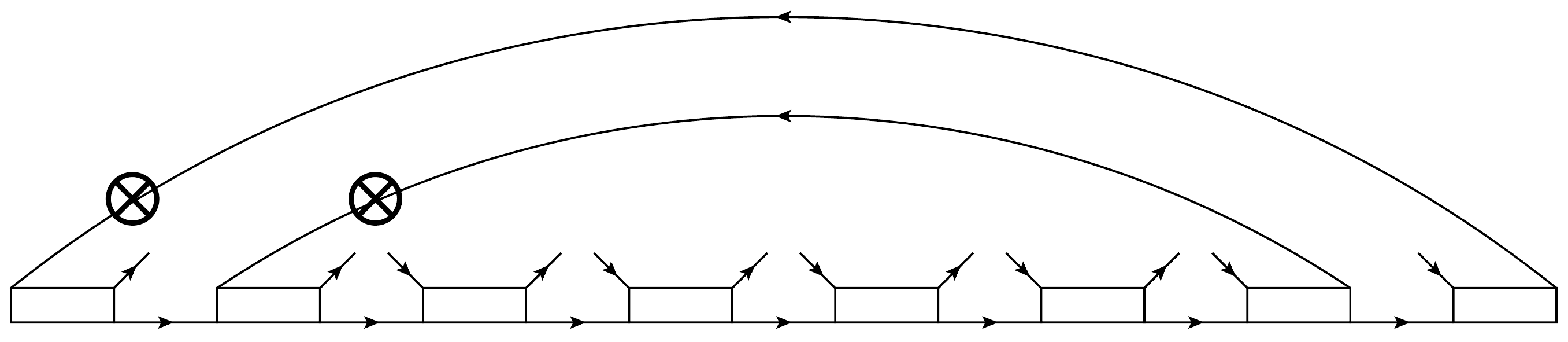}
\caption{Basic structure of a class of self-energy diagrams that contributes to the asymptotic large-order behavior given in Eq.~(\ref{eq:RNF}).  
The diagrams are formed by closing the open ends with $G^0_{\uparrow}$-propagators. Besides, we also consider diagrams obtained by 
interchanging the two $G^0_{\uparrow}$-propagators marked with the symbol $\otimes$.
}
\label{fig:twobackward}
\end{figure}

Next we consider the diagrams based on the structure shown in Fig.~\ref{fig:twobackward}.  There are two backward spin-up propagators which are shown and, like before, whose ends can be interchanged. First we consider the class where all the open ends are connected by either (i) forward spin-up propagators, or (ii) backward spin-up propagators closing a single $\Gamma^0$-line. 
A second class is obtained by allowing all possible connections of the open spin-up ends. 
Note that such restrictions on the topology are easy to implement in the current algorithm, since they can be achieved by setting the right matrix elements to zero before calculating the determinant. Both cases revealed yet two other functions $F_{\rm class}(\tau) \neq F_{\rm all}(\tau)$, while still having the same value of $R$.

\begin{figure}
\includegraphics[scale=0.38,width=0.79\columnwidth]{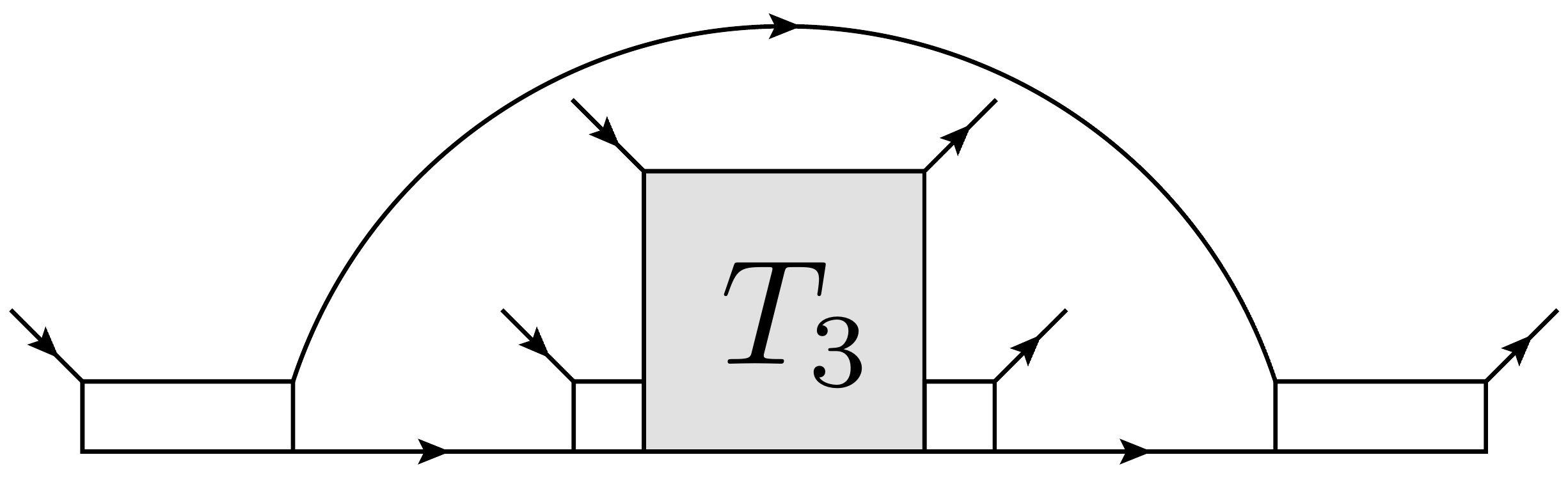}
\caption{Basic structure of a class of self-energy diagrams that contributes to the asymptotic large-order behavior given in Eq.~(\ref{eq:RNF}).  
One obtains six self-energy diagrams by closing the open ends with $G^0_{\uparrow}$-propagators. 
}
\label{fig:T3closed}
\end{figure}

Finally, we consider the six diagrams based on the structure shown in Fig.~\ref{fig:T3closed}. 
The structure of the three-body propagator $T_3$ is shown in Fig.~\ref{fig:T3diagrams}. 
The six diagrams are obtained by closing the open ends with $G^0_{\uparrow}$-propagators in all possible ways. 
Each of these six diagrams  gives a different $F_{\rm diag}(\tau)$ which contributes to $F_{\rm all}(\tau)$, while their sum is much smaller than the leading 
behavior of $F_{\rm all}(\tau)$. 
It is easy to come up with more topologies, similar to the ones considered in this appendix, that contribute to $F_{\rm all}(\tau)$ in 
a significant way.

\bibliography{felix_copy.bib}

\end{document}